\documentclass[journal=jacsat,manuscript=article]{achemso}

\usepackage[version=3]{mhchem}
\usepackage{textcomp,mathcomp}
\usepackage{gensymb}

\author{Tanya Berry}
\affiliation{Department of Chemistry, Princeton University, Princeton, NJ 08540}
\author{Jaime M. Moya}
\affiliation{Department of Chemistry, Princeton University, Princeton, NJ 08540}
\author{David Smiadak}
\affiliation{Department of Chemical Engineering and Material Science, Michigan State University, East Lansing, MI 48824}
\author{Scott B. Lee}
\affiliation{Department of Chemistry, Princeton University, Princeton, NJ 08540}
\author{Sigalit Aharon}
\affiliation{Department of Chemistry, Princeton University, Princeton, NJ 08540} 
\author{Alexandra Zevalkink}
\affiliation{Department of Chemical Engineering and Material Science, Michigan State University, East Lansing, MI 48824}
\author{Tyrel M. McQueen}
\affiliation{Department of Chemistry, Department of Materials Science and Engineering, Institute for Quantum Matter, William H. Miller III Department of Physics and Astronomy, The Johns Hopkins University, Baltimore, MD 21218}
\alsoaffiliation{Department of Chemistry, The Johns Hopkins University, Baltimore, MD 21218}
\alsoaffiliation{Department of Materials Science and Engineering, The Johns Hopkins University, Baltimore, MD 21218}
\author{Leslie M. Schoop}
\email{lschoop@princeton.edu}
\affiliation{Department of Chemistry, Princeton University, Princeton, NJ 08540}

\title
  {Bonding Interactions Can Drive Topological Phase Transitions in a Zintl Antiferromagnetic Insulator}

\begin{document}

\begin{abstract}

While  $\sim$30\% of materials are reported to be topological, topological insulators are rare. Magnetic topological insulators (MTI) are even harder to find. Identifying crystallographic features that can host the coexistence of a topological insulating phase with magnetic order is vital for finding intrinsic MTI materials. Thus far, most materials that are investigated for the determination of an MTI are some combination of known topological insulators with a magnetic ion such as MnBi$_2$Te$_4$. Motivated by the recent success of EuIn\textsubscript{2}As\textsubscript{2}, we investigate the role of chemical pressure on topologically trivial insulator, Eu$_5$In$_2$Sb$_6$  via Ga substitution. Eu$_5$Ga$_2$Sb$_6$  is predicted to be topological but is synthetically difficult to stabilize. We look into the intermediate compositions between Eu$_5$In$_2$Sb$_6$ and Eu$_5$Ga$_2$Sb$_6$ through theoretical works to explore a topological phase transition and band inversion mechanism. We attribute the band inversion mechanism to changes in Eu-Sb hybridization as Ga is substituted for In due to chemical pressure. We also synthesize Eu\textsubscript{5}In\textsubscript{4/3}Ga\textsubscript{2/3}Sb\textsubscript{6}, the highest Ga concentration in Eu\textsubscript{5}In\textsubscript{2-x}Ga\textsubscript{x}Sb\textsubscript{6}, and report the thermodynamic, magnetic, transport, and Hall properties. Overall, our work paints a picture of a possible MTI via band engineering and explains why Eu-based Zintl compounds are suitable for the co-existence of magnetism and topology.
\end{abstract}

\section{Introduction}
 Since the early 2000s with the Kane-Mele model, there has been a significant advancement not only in the theoretical classification of topological materials but also in their experimental realization. 
\cite{kane2005quantum,zhang2009topological, liu2014discovery, bradlyn2016beyond, schoop2016dirac, deng2020quantum, riberolles2021magnetic} 
 There have also been advancements in discovering new topological materials with a multi-pronged approach that includes computation, chemical design principles, and data-driven approaches.{\cite{regnault2022catalogue, vergniory2019complete, national2009frontiers,schoop2018chemical} Still, when it comes to the experimental realization of topological insulators (TI's), few ideal candidates are known, such as HgTe, monolayer WTe$_2$, Bi\textsubscript{1-x}Sb\textsubscript{x}, Bi$_2$Te$_3$, Bi$_2$Se$_3$, and variations of those. These TI's exhibit various phenomena, such as the quantum spin Hall effect, and are well studied.\cite{hasan2010colloquium, cava2013crystal, chen2009experimental,konig2007quantum, bernevig2006quantum, teo2008surface, csahin2015tunable, qi2010quantum} } 
 TIs are insulators that undergo band inversion and a change in parity. These insulators are unique from regular insulators; while being insulating in the bulk, they feature metallic, spin-momentum-locked, topological surface states (TSSs). To increase the library of materials that show band inversion, which is crucial for topological insulators, the scientific community has been searching for design principles that assist in rationally affecting the band structure of a material by playing with its' elemental building blocks. \cite{khoury2021chemical, schoop2018chemical, zhu2012band, kumar2020topological}

Several band inversion mechanisms for non-magnetic TIs have been discussed in the literature, including spin-orbit coupling (SOC), the inert pair effect (IPE), strain, band folding, and orbital hybridization. \cite{yan2010theoretical, zhu2012band, fu2007topological, zhang2009topological, zhang2019topological, yan2013large, jiang2023flat, kumar2020topological, isaeva2013bismuth, yan2015topological, bernevig2006quantum, konig2007quantum, chadov2010tunable, wang2022highly, zhao2015strain, huang2013nontrivial, weng2014transition, schoop2016dirac, honma2023unusual, ren2015single, kooi2020chalcogenides, cuono2023ab, yan2012topological, cao2020trivial, fuhrman2015interaction} 
The early TI's were mostly SOC and IPE-driven. For example, in Bi$_2$Te$_3$-type materials, the band inversion only appears when SOC is included in the calculation and without SOC there is no inversion. \cite{zhang2009topological} 
Accounting for SOC does two things. It a) inverts the bands and b) opens a gap in the now inverted bands.\cite{muchler2013topological} 
In the case of IPE-driven TI's, the band inversion appears due to the low lying 6$s$ orbital (the inert pair) and thus appears mostly in TIs made of heavy $p$-block elements.\cite{kumar2020topological, isaeva2013bismuth, yan2015topological, bernevig2006quantum, konig2007quantum, chadov2010tunable} 
HgTe, the first known TI, is an example of the IPE-driven mechanism.
When materials are synthesized as thin films, strain can be more easily applied, which allows for strain-driven band inversion, as seen in Bi (111) films. \cite{huang2013nontrivial} 
If the materials crystallize in a non-symmorphic space group, the bands will be folded due to the additional translational symmetry. Band folding is a convenient way to introduce band inversions and is the reason bands are inverted in elemental Bi or square-net materials such as ZrSiS and Bi.\cite{schoop2016dirac, khoury2021chemical,schindler2018higher} 
Finally, band inversions can appear due to the hybridization of e.g. \textit{f} and \textit{d} bands. Such hybridization effects are seen in SmB\textsubscript{6}.\cite{fuhrman2015interaction} 

No matter how the band inversion is induced, TI's become even more interesting when they exhibit magnetic order. Magnetic order and the consequential breaking of time-reversal symmetry, can open a gap in the TSSs of a TI, and enable the realization of the quantum anomalous Hall effect (QAHE), which enables dissipation-less transport along edges of a 2D TI.\cite{tokura2019magnetic} 
Such a material is known as a magnetic topological insulator (MTI), which is subdivided into two categories: ferromagnetic and antiferromagnetic. Ferromagnetic topological insulators (FM TI's) can arise from ferromagnetic doping, in the case of Cr-doped (Bi,Sb)$_2$Te$_3$, which exhibits the QAHE at 30 mK,\cite{chang2013experimental} 
or from intrinsically ferromagnetic topological insulators, such as MnBi$_{2}$Te$_{4}$ thin films, which show QAHE depending on the number of layers.\cite{li2019intrinsic} 
The second category consists of antiferromagnetically ordered topological insulators (AFM TIs), which divides further into two subcategories; first are intrinsic antiferromagnetic topological insulators, also referred to as Axion insulators, where the antiferromagnetic order gaps-out all of the surface states (as also seen in antiferromagnetic MnBi$_{2}$Te$_{4}$ thin films or single crystals).\cite{yan2019crystal, hao2019gapless} 
In such materials a quantized magnetoelectric response is expected.\cite{wu2016quantized} 
Second are generalized antiferromagnetic topological insulators, where only a subset of the surface states are gapped. This phenomenon has been observed in many recent studies of Eu-based Zintl materials, such as EuIn$_{2}$As$_{2}$, Eu$_{3}$In$_{2}$As$_{4}$ and EuCd$_{2}$As$_{2}$.\cite{xu2019higher, zhao2024hybrid, ma2020emergence} 
Designing a magnetic topological insulator introduces an additional complication in that the magnetic states cannot interfere with the band inversions giving rise to topology. This is why most magnetic topological insulators are derived from known non-magnetic TIs with the addition of a magnetic constituent, as in MnBi$_2$Te$_4$. We have previously identified that Zintl compounds are a promising material family to host magnetic topological insulators.\cite{varnava2022engineering} 
The first design consideration is per the Zintl formalism: there are ionic (salt-like) interactions between cations and polyanionic networks, producing a large energy separation between cations and anions frameworks. This large energy separation provides autonomous chemical handles to tune the magnetism and topology, independently. The second design consideration is that, when looking at cations, we want a magnetically isotropic ion such that it can be polarized easily with an applied magnetic field. It should also be compatible with the oxidation states commonly found in Zintl materials. Since a majority of the Zintl cations have an oxidation state of 2+, Eu$^2$$^+$ ($J$=$S$=7/2 and $L$=0) is an ideal choice. The third design consideration is that, when considering polyanionic frameworks, we want it to be composed of heavy atoms that have small differences in electronegativity, so they have a small band gap and also a large spin-orbit coupling (SOC), to potentially induce a band gap in an inverted band structure.

Eu\textsubscript{5}In\textsubscript{2}Sb\textsubscript{6} meets all of the design criteria with [Eu\textsubscript{5}]\textsuperscript{10+} as the magnetic cation framework and [In\textsubscript{2}Sb\textsubscript{6}]\textsuperscript{10-} as the polyanionic framework (i.e. 2 In\textsuperscript{3+}, 4 Sb\textsuperscript{3-}, and 1 (Sb\textsubscript{2})\textsuperscript{4-}). Eu$_5$In$_2$Sb$_6$ is a trivial magnetic insulator.\cite{rosa2020colossal,varnava2022engineering}
It was predicted with density functional theory (DFT) that an isoelectronic substitution of Ga in the In position would lower the lattice constant. This, in turn, can drive a band inversion, and make it topological.\cite{varnava2022engineering, morano2024noncollinear,zevalkink2012influence} 
Naively, one might not expect a topological inversion, given the lower SOC in Ga or In. However, one might view the substitution of Ga for In as a chemical pressure effect that increases the bandwidth and this allows the gap to close and invert.\cite{chanakian2015high} 
Density Functional Theory is used to identify band inversion in materials and to better understand the mechanism that drives topological surface states through band inversion. However, DFT alone does not provide a full chemical understanding of the origin of such inversions, limiting our ability to create new materials that harbor such inversions. 

Here, we show that the exact band inversion mechanism in Eu\textsubscript{5}In\textsubscript{2-x}Ga\textsubscript{x}Sb\textsubscript{6} is more complex, involving not only the anionic framework but also cationic states, a ``third party'' bonding effect. This mechanism has not been previously explored to our knowledge and provides a new route to the creation of magnetic topological insulators. Further, we synthesize Eu\textsubscript{5}In\textsubscript{2-x}Ga\textsubscript{x}Sb\textsubscript{6} to a maximal achievable substitution of x=$\frac{2}{3}$, and report the magnetic and electronic properties of the material in single crystal form, showing that the trends are consistent with our calculations. We will help in the theoretical and practical design of future MTIs, and the broadening of the materials library.

\subsection{Results and Discussion}
\subsection{Theoretical Investigation of Band Inversion}
To investigate the nature of the band inversion from topologically trivial, Eu\textsubscript{5}In\textsubscript{2}Sb\textsubscript{6} with ${Z}$$_2$ = 0 to topologically non-trivial, hypothetical Eu\textsubscript{5}Ga\textsubscript{2}Sb\textsubscript{6} (${Z}$$_2$ = 1), we first perform PBE+SOC (Perdew-Burke-Ernzerhof) calculations on a series of compositions Eu\textsubscript{5}In\textsubscript{2-x}Ga\textsubscript{x}Sb\textsubscript{6} for $0 \leq x \leq 2$. Since the topology is sensitive to the crystal symmetry, as well as the number of bands at time-reversal invariant momenta, we carried out the calculations on ordered 1x1x3 supercells of the crystallographic unit cell of Eu\textsubscript{5}In\textsubscript{2}Sb\textsubscript{6} containing mixtures of In and Ga. The 3-fold supercell was chosen because it is the smallest supercell that retains ${Pbam}$ symmetry, it is capable of expressing In/Ga mixtures at relevant compositions (x= 0, $\frac{2}{3}$, $\frac{4}{3}$, and 2), and the 3-fold enlargement will preserve the parity of bands at the $\Gamma$ point, where the band inversion was previously identified (-1\textsuperscript{3} = -1). The PBE+SOC relaxed unit cells of Eu\textsubscript{5}In\textsubscript{2-x}Ga\textsubscript{x}Sb\textsubscript{6} (x= 0, $\frac{2}{3}$, $\frac{4}{3}$, and 2) are given in Table S1.

\begin{figure}
  \includegraphics[scale=0.45]{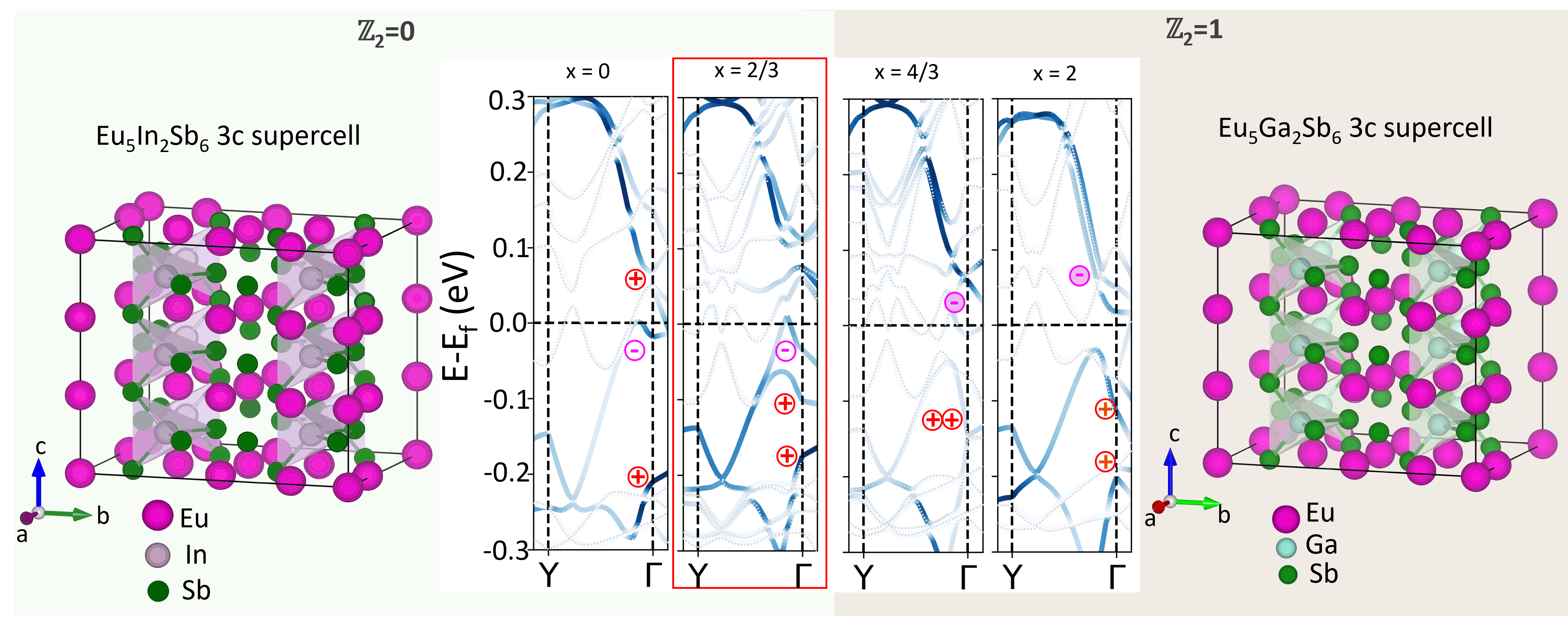}
  \caption{Band structures (PBE+SOC)of the Eu\textsubscript{5}In\textsubscript{2-x}Ga\textsubscript{x}Sb\textsubscript{6} solid solution for (a) x=0, (b) x=$\frac{2}{3}$, (c) x=$\frac{4}{3}$, and (d) x=2, from $Y$ (0,$\frac{1}{2},0$) to $\Gamma$ ($0,0,0$). The intensity of the dots indicates the fractional contribution of that band to the average structure. The composition discussed later in this work is Eu\textsubscript{5}In\textsubscript{2-x}Ga\textsubscript{x}Sb\textsubscript{6} with x=$\frac{2}{3}$ which is also squared in red. The parities of the $Y$ to  $\Gamma$ point indicate a change in between x=$\frac{2}{3}$ and x=$\frac{4}{3}$, which implies a change in band inversion as a function of Ga substitution in the Eu\textsubscript{5}In\textsubscript{2-x}Ga\textsubscript{x}Sb\textsubscript{6} structure. At x=$\frac{4}{3}$ and 2 in the  Eu\textsubscript{5}In\textsubscript{2-x}Ga\textsubscript{x}Sb\textsubscript{6} solid solution the purple circled minus sign is above the Fermi level illustrating the switch in the parity that derives the topological phase tranistion which is not the case in the In-rich compositions.  The structures of the 3c supercell of Eu$_5$In$_2$Sb$_6$ and Eu$_5$Ga$_2$Sb$_6$ are generated from the PBE+SOC calculations. It is important to note that these are not the actual crystal structures of the materials. } 
  \label{bands}
\end{figure}
Band structure plots unfolded to the crystallographic Eu\textsubscript{5}In\textsubscript{2}Sb\textsubscript{6} unit cell are shown in Figure 1 and Figure S1. When moving from x=0 to x=2, we observe significant band shifts at the $\Gamma$ point. The lowest lying, positive parity, conduction band state above the Fermi level at x=0 moves downward in energy to below $E$$_f$ at x$\geq$$\frac{2}{3}$. The highest lying, negative parity, valence band state below the Fermi level at x=0 first moves slightly down in energy at x=$\frac{2}{3}$ before increasing in energy and crossing above $E$$_f$ at x=$\frac{4}{3}$. It is the movement of this second band that causes the change in topology. These trends are in agreement with the previous calculation at x=0 and x=2 end members.

To determine where the topological phase transition occurs, the ${Z}$$_2$ index was calculated for each composition. Inversion symmetric insulators are ${Z}$$_2$ nontrivial if 
\begin{equation*}
  \prod_{n,m=0.1} \prod_{\alpha \in filled } \xi ^{(\alpha)}_{nm}=-1
\end{equation*}
where $\xi ^{(\alpha)}_{nm}$ are the inversion eigenvalues corresponding to the $\alpha$th Kramers doublet of bands at the momenta  $\Gamma_{nm}$=($n\kappa_1$ + $m\kappa_2$)/2 ($n,m =0,1$) and $\kappa_i$ are the reciprocal lattice vectors.\cite{fu2007topological} Utilizing the cross products at different time-reversal invariant momenta (TRIM) points as listed in Table 1, we determine that the change in parity happens between x=$\frac{2}{3}$ and $\frac{4}{3}$ in the Eu$_5$In$_{2-x}$Ga$_x$Sb$_6$ system.

However, since the generalized gradient approximation (GGA), to which the PBE functional belongs, usually tends to underestimate band gaps and in this specific case it predicts negative band gaps i.e. metallic band structures, which is in contrast to experiments. We also carried out band structure calculations using the modified Becke-Johnson (mBJ) functional+SOC and the relaxed atomic coordinates and unit cells from the PBE+SOC calculation, Table~S2. 
In contrast with PBE+SOC, mBJ+SOC predicts positive band gaps for all compositions, Figure-S2.

The magnitude of the predicted band gap is $\approx$55 meV which is in reasonable agreement with the transport gap.\cite{rosa2020colossal} Carrying out a parity analysis in  Table-S2, shows that the x=0, $\frac{2}{3}$, and $\frac{4}{3}$ compositions are topologically trivial and x=2 is topological non-trivial. As this result qualitatively agrees with the PBE (i.e both calculations confirm a band inversion appears eventually upon Ga substitution), we will proceed with PBE calculations to further investigate the mechanistic details of the inversion.
We now turn to the chemical origin of this topological phase transition. Intuitively one might think that lowering the SOC upon substituting Ga for In is disadvantageous for creating a topological insulator. However, it should be kept in mind that SOC is what gaps inverted bands and only rarely induces band inversions. To deduce a chemical picture of why the substitution of In by Ga results in a topological band inversion, we carried out a Crystal Orbital Hamilton Population (COHP) analysis, Figure 2 and Figure S3.

\begin{table}
  \label{parity}
  \centering
  \begin{tabular}{ccccccccc}
    \hline
          x ratios&$\Gamma$&          $X$&$Y$&$Z$&$U$&$T$&$S$&  $R$\\
    \hline
          x=0&1&     1&1&1&1&1&1&  1\\
          x=$\frac23$&1&     1&1&1&1&1&1&  1\\
          x=$\frac43$&-1&     1&1&1&1&1&1&  1\\
          x=2&-1&     1&1&1&1&1&1&  1\\\hline
  \end{tabular}
  \caption{The parity outputs at the k points using GGA+SOC in Eu\textsubscript{5}In\textsubscript{2-x}Ga\textsubscript{x}Sb\textsubscript{6}.}
\end{table}

The COHP method provides an energy-weighted picture of the bonding of a solid; negative (positive) numbers indicate the energy gained (lost) associated with bonding (antibonding) states respectively. In the COHP, averaged over all Eu-Sb, In-Sb, and Ga-Sb atom pairs, the values are largely unchanged across the series, except around -4 to -6 eV, and close to the Fermi level between, -1 to 0 eV (as seen in the inset of Figure 2). The changes in the former are reflective of changes in bonding in the cationic or anionic frameworks, while the latter reflect the changes associated with the band motions driving the topological phase transition. 

We can gain further insights into the local bonding responsible for each of these features by looking at subsets of bonded pairs. In-Sb and Ga-Sb bonding forms the anionic framework of the structure. The pCOHP for these bonds show large changes in the -4 to -6 eV region but are all virtually superimposable around 0 eV (Figure 2(b) inset). Integration of the pCOHP of In/Ga-Sb in the energy window of 0 to -6 eV, where valence shell bonding interactions are expected to occur, reveals increasing Ga substitution disfavors the anionic structure from x=0 (-94.953), through x=$\frac{2}{3}$ (-92.235) and x=$\frac{4}{3}$ (-89.261) to x=2 (-86.394). This result is consistent with the experimental observation that phases with increased Ga substitution can not be synthesized. The observed similarities between these pCOHP curves near the Fermi energy, however, implies that while In-Sb and Ga-Sb bonding is crucial to the stability of the anionic framework, it is not (directly) responsible for the change in topology. In contrast, the pCOHP for Eu-Sb bonds shows smaller changes from -1 to -6 eV, but much more marked changes in the region near 0 eV, associated with the band motions that change the topology of the compound. This demonstrates that it is changes in Eu-Sb bonding (and, in particular, a stabilization of Eu-Sb bonding) that drive the topological phase transition. This explains why substituting Ga can produce a topological state: it destabilizes bonding within the anionic network; this, in turn, provides a driving force towards the formation of Eu-Sb bonds (which involve formal charge transfer from Eu to the anionic framework).
\begin{figure}
    \centering
    \includegraphics[width=0.61\linewidth]{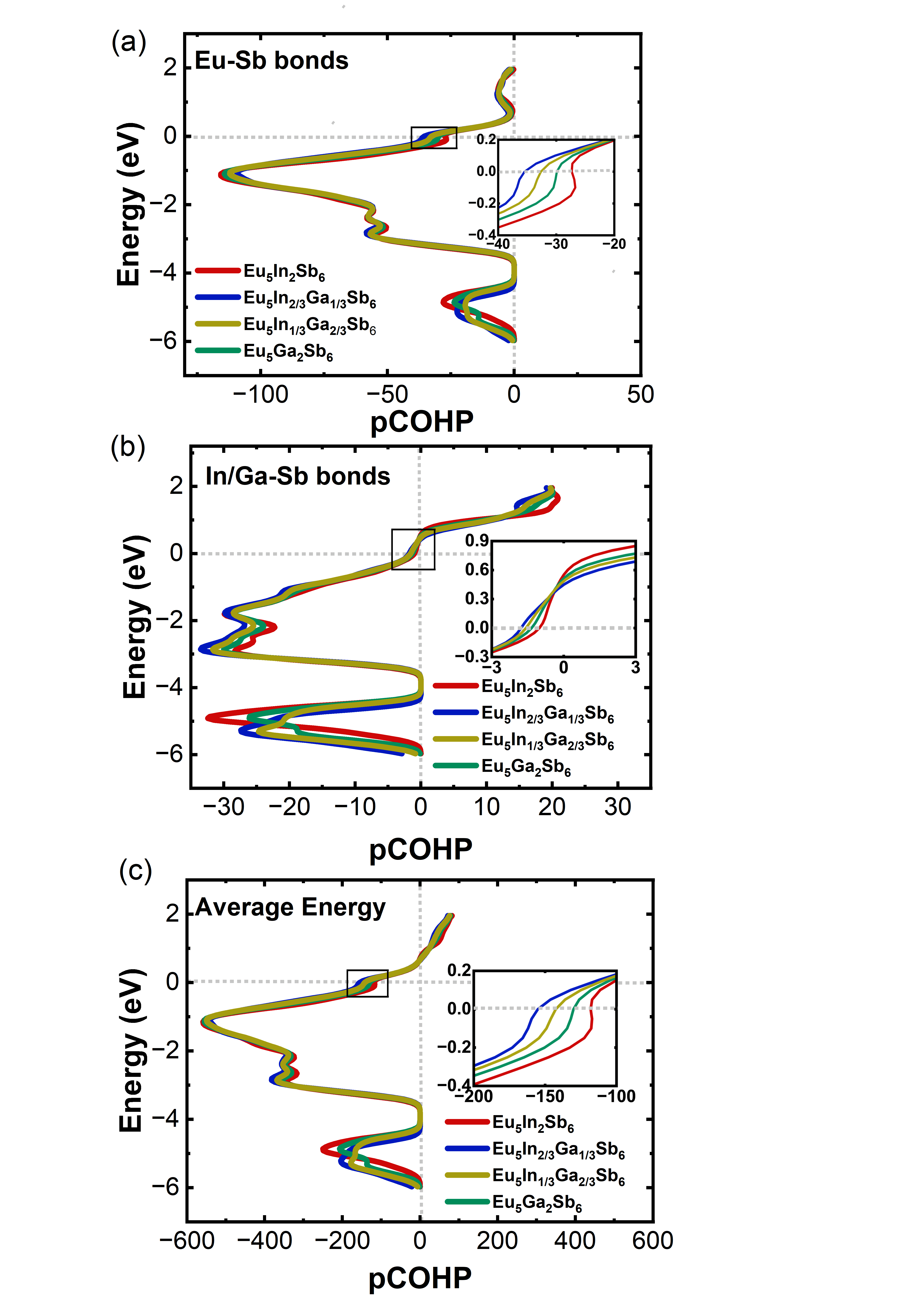}
    \caption{The integrated pCOHP/bond analysis for (a) the average, (b) In/Ga-Sb bonds, and (c) Eu-Sb bonds for the Eu\textsubscript{5}In\textsubscript{2-x}Ga\textsubscript{x}Sb\textsubscript{6} structure in x=0, x=$\frac{1}{3}$, x=$\frac{2}{3}$, and x=1 respectively. The inset shows zoomed in views of the frames boxed in the three panels.}
    \label{cohp}
\end{figure}
To our knowledge, this “third party effect” or complex hybridization has not been previously considered in the design of magnetic topological insulators in Zintl compounds. In most cases, hybridization effects are seen in a variety of materials such as EuCd\textsubscript{2}As\textsubscript{2}, CeOs\textsubscript{4}P\textsubscript{12}, CeOs\textsubscript{4}As\textsubscript{12}, Ce\textsubscript{3}Pt\textsubscript{3}Bi\textsubscript{4}, Ce\textsubscript{3}Pd\textsubscript{3}Bi\textsubscript{4}, Ta\textsubscript{4}SiTe\textsubscript{4}, and SmB\textsubscript{6}.\cite{cuono2023ab, yan2012topological,cao2020trivial, fuhrman2015interaction} 
However, in most of the cases there is a coupling of $f$ and $d$ states.  Recent works on GdTe$_2$ and EuTe$_2$ have shown that complex $p$-$d$-$f$ hybridization can result in purely $p$-based physics of topological transport. \cite{zeer2024promoting} In our work, we see a coupling of Eu-$d$ and Ga(In)/Sb-$p$ states that drive the topological phase transition. Further Zintl-MTIs can likely be discovered through careful application of this “third party effect". 

\subsection{Single Crystal Growth and Structural Characterization of Eu\textsubscript{5}In\textsubscript{2-x}Ga\textsubscript{x}Sb\textsubscript{6} }
With this chemical understanding in mind, it is important to ask to what degree these predictions are in agreement with experimental observations. The synthesis objective is to achieve Ga isoelectronic substitution in the In site in  Eu\textsubscript{5}In\textsubscript{2-x}Ga\textsubscript{x}Sb\textsubscript{6} as x$\rightarrow$2 (near the Ga-rich phase), while maintaining the ${Pbam}$ space group symmetry. To investigate the threshold of Ga substitution in the Eu\textsubscript{5}In\textsubscript{2-x}Ga\textsubscript{x}Sb\textsubscript{6} structure, a series of synthesis attempts were carried out using flux, traditional solid synthesis, and chemical vapor transport methods. Of these synthesis approaches flux synthesis was most successful. These synthetic approaches were also employed to synthesize the hypothetical Eu\textsubscript{5}Ga\textsubscript{2}Sb\textsubscript{6} (these form as minority phases that were not reproducible), however the phase was not stabilized and we formed the thermodynamically stable phase of EuGa$_2$Sb$_2$ instead.\cite{berry2021antiferro} 
\begin{figure}
    \centering
    \includegraphics[width=0.85\linewidth]{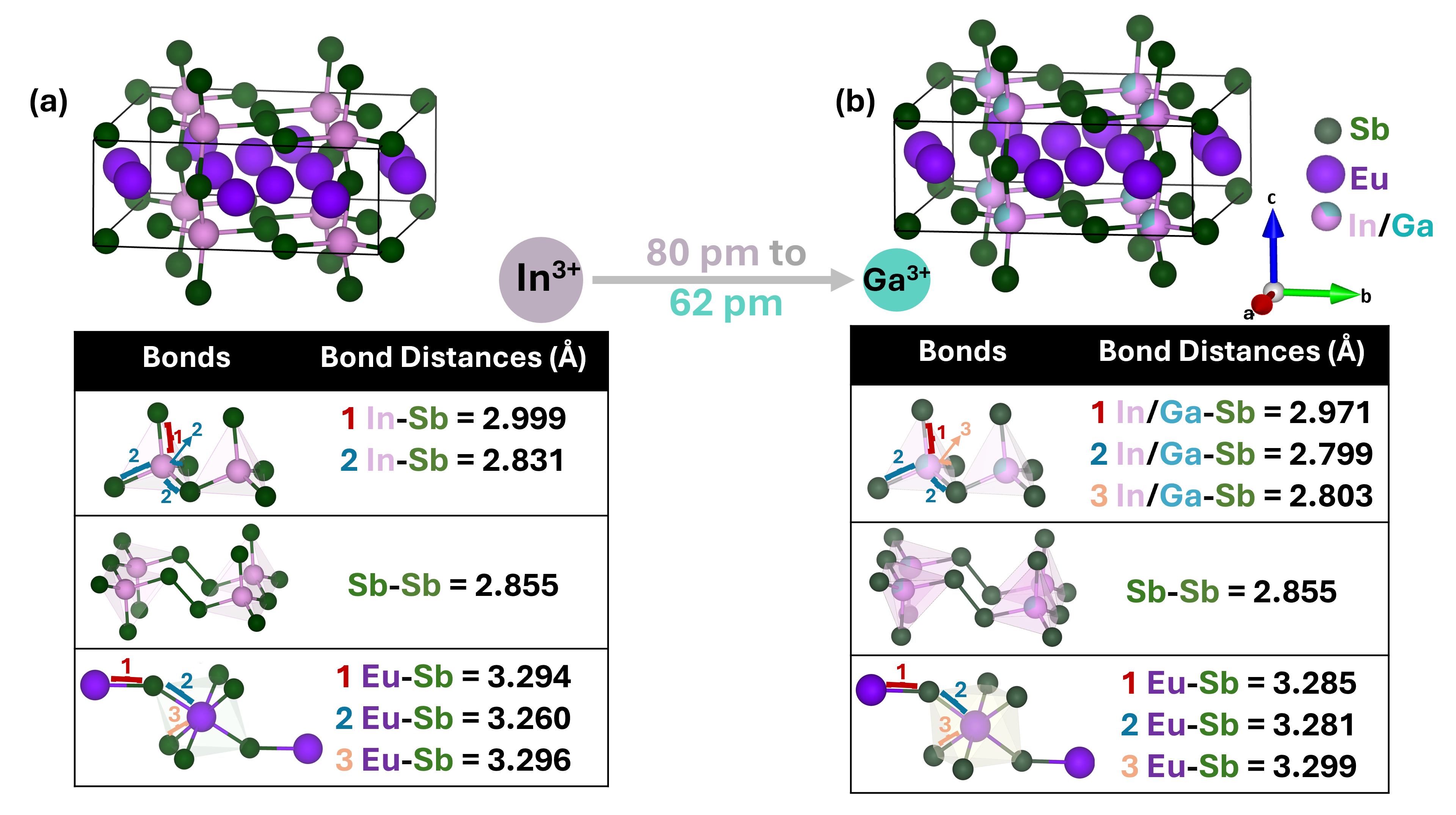}
    \caption{Shift structural shifts from (a) Eu\textsubscript{5}In\textsubscript{2}Sb\textsubscript{6} to (b) x=$\frac{2}{3}$ Eu\textsubscript{5}In\textsubscript{2-x}Ga\textsubscript{x}Sb\textsubscript{6}. The tables below (a) and (b) explain the distortion in the trigonal pyramidal to a bent geometry in the (InSb$_4$) frameworks and the overall shortening of the bond distances on Ga substitution. Please see Figure-S6 for more significant figures and errors of the bond lengths given.}
    \label{struc}
\end{figure}

\begin{figure}
\includegraphics[scale=0.165]{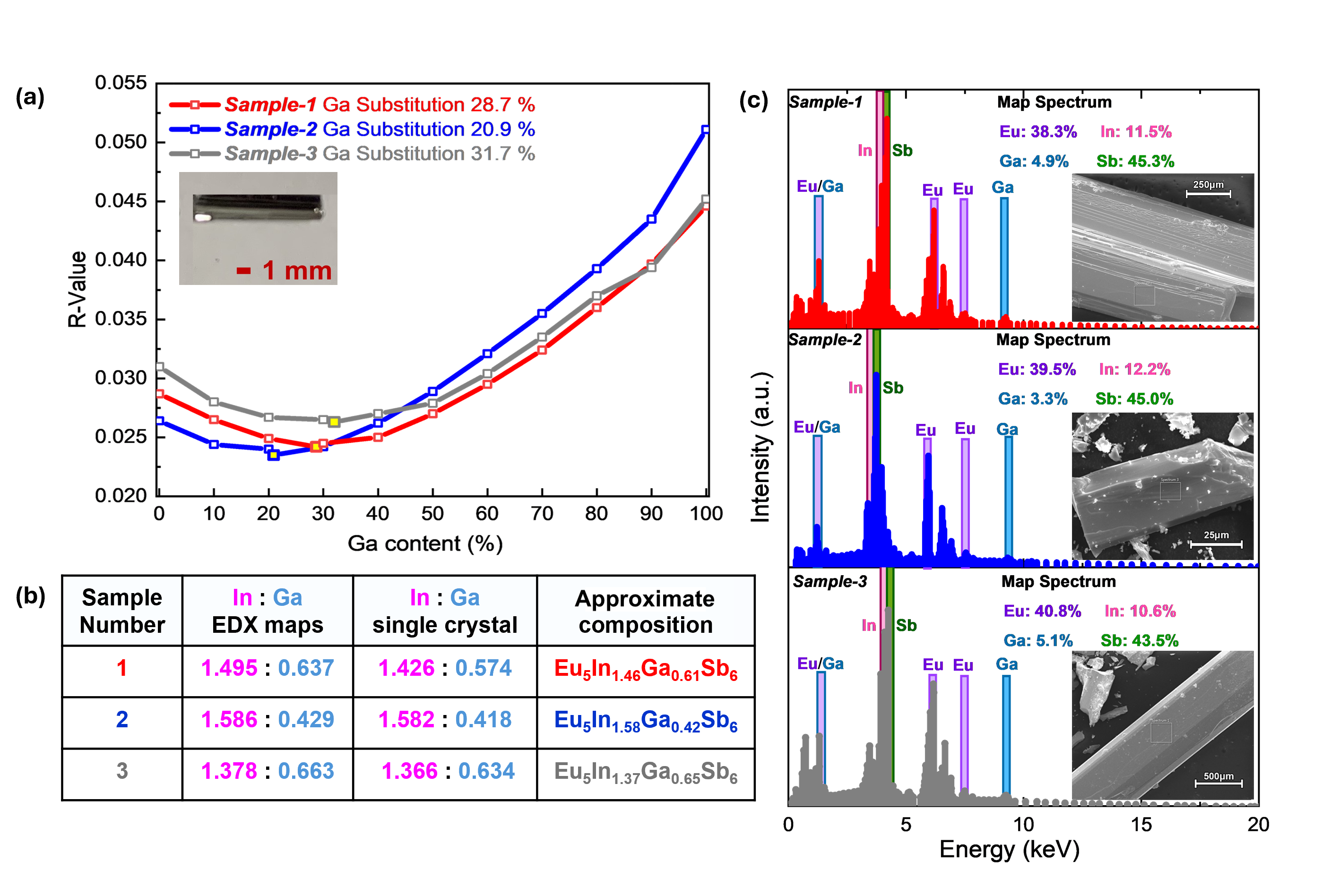}
\caption{(a) R-value and Ga content in different Ga samples. These results show that different Ga substations are possible in the series. The single-crystal solutions to these are in the SI. The data points with yellow highlights are the ones with the lowest R-value. The inset shows A picture of the single crystal of Eu\textsubscript{5}In\textsubscript{2-x}Ga\textsubscript{x}Sb\textsubscript{6} with x$\approx$$\frac{2}{3}$. The inset show a picture of the single crystal of Eu\textsubscript{5}In\textsubscript{2-x}Ga\textsubscript{x}Sb\textsubscript{6} with x$\approx$$\frac{2}{3}$. (b) The overall composition determined by SEM-EDX and single crystal stricture solutions. (c) The SEM-EDX maps of a few different compositions of Eu\textsubscript{5}In\textsubscript{2-x}Ga\textsubscript{x}Sb\textsubscript{6}. The inset shows the SEM of a few of the single crystals.}
\label{syn}
\end{figure}

As seen in Figure 3, the In$^{3+}$ and Ga$^{3+}$ ionic radii are 80 pm and 62 pm which are quite different.\cite{shannon1968synthesis, shannon1970effective} 
The maximum x synthetically achievable in the Eu\textsubscript{5}In\textsubscript{2-x}Ga\textsubscript{x}Sb\textsubscript{6} series was x$\approx$$\frac{2}{3}$. The determination of the x$\approx$$\frac{2}{3}$ was done using single crystal X-ray diffraction (SXRD) utilizing the R-value analysis and scanning electron microscopy-energy dispersive X-ray analysis (SEM-EDX) difference in the In:Ga contents, as seen in Figure 4 and Table S3-S5.

The difference in form factors between the two species causes local packing tensions effects and notable shifts in the bond lengths also shown in Figure 3. If we consider the crystal structures of Eu\textsubscript{5}In\textsubscript{2}Sb\textsubscript{6}, it becomes evident that the distortion of the polyhedra becomes most prominent as we approach the x$\approx$$\frac{2}{3}$ limit in the Eu\textsubscript{5}In\textsubscript{2-x}Ga\textsubscript{x}Sb\textsubscript{6} structure. This distortion suggests the collapse of the structure upon a higher threshold of Ga substitution in the ${Pbam}$ structure.

The Ga to InSb ratios derived from SXRD and EDX characterization do not match the nominal ratios given by the flux synthesis i.e. the input Ga concentration is not monotonically related to InSb flux ratios. Such deviations in flux synthesis are common and suggest that there is a set concentration where the flux composition reaches a dissolving minimum as a function of temperature. However, within each batch various crystals were compared and the composition was consistent as seen via repeated flux growths and SXRD studies. Previous work aiming to substitute Cd and Zn in the Eu\textsubscript{5}In\textsubscript{2}Sb\textsubscript{6} system found an even lower threshold of x less than 0.1.\cite{lv2017cd}\cite{park2002eu, chanakian2015high}
Overall, a composition of x$\approx$$\frac{2}{3}$ in Eu\textsubscript{5}In\textsubscript{2-x}Ga\textsubscript{x}Sb\textsubscript{6} was achieved in the single crystal form and it is close to the topological phase transition.

\subsection{Transport Behavior of Eu\textsubscript{5}In\textsubscript{4/3}Ga\textsubscript{2/3}Sb\textsubscript{6}}
Next, we study the magnetic and electronic properties of the compound with the highest Ga content we could grow, Eu\textsubscript{5}In\textsubscript{4/3}Ga\textsubscript{2/3}Sb\textsubscript{6}. While at this substitution level, the topological phase transition is not yet expected to occur, it is still of interest to study how achievable even this level of substitution affects the properties as compared to Eu$_5$In$_2$Sb$_6$, which has been shown to have a non-collinear magnetic structure and colossal magnetoresistance.\cite{rosa2020colossal,morano2024noncollinear} 
In our Ga-substituted samples, we observe that the antiferromagnetic transition temperatures arise from Eu\textsuperscript{2+} \textit{S}=\textit{J}=7/2 to have similar features to non-Ga substituted Eu$_5$In$_2$Sb$_6$.\cite{rosa2020colossal,morano2024noncollinear}
A discussion of the magnetization as a function of field and temperature as well as heat capacity results explaining the nature of antiferromagnetism in Eu\textsubscript{5}In\textsubscript{4/3}Ga\textsubscript{2/3}Sb\textsubscript{6} is described in the supplemental text and Figure-S4, S5 respectively.

To see the implication of chemical pressure on the electronic properties of Eu\textsubscript{5}In\textsubscript{4/3}Ga\textsubscript{2/3}Sb\textsubscript{6} single crystals, resistivity as a function of temperature with various fields was investigated.\textbf{ }Figure 5 shows the resistivity was measured $\mu_{o}\textit{H}$//b and j$\perp$b. At \textit{T} $<$15 K we observe three transitions that overlap with the magnetic transition temperatures as observed in the \textit{M}(\textit{T}) plots at $\mu_{o}\textit{H}$//b at $\mu_{o}\textit{H}$=0.1T. Therefore, those features in the resistivity could be attributed to the loss of spin disorder scattering. However, at \textit{T}$ \sim$60 K, there is a change in the resistivity i.e. \textit{T}$<$60 K the resistance of Eu\textsubscript{5}In\textsubscript{2-x}Ga\textsubscript{x}Sb\textsubscript{6} with  x=$\frac{2}{3}$ behaves as a metal (d$\rho/\text{d}T>0$), and \textit{T}$>$60 K it behaves as an insulator (d$\rho/\text{d}T<0$). The feature at \textit{T}$ \sim$60 K can be decoupled to the magnetic order since it is above the ordering temperatures of Eu. Such a feature is commonly seen in topological insulators such as Bi\textsubscript{2}Se\textsubscript{3} and related materials and is often attributed to changes in electronic surface states or defects.\cite{kushwaha2016sn} On applying $\mu_{o}\textit{H}$ of up to 9 T both the high-temperature and magnetic anomalies are suppressed. This suppression suggests a strong coupling that occurs with a combination of effects such as suppression of strong magnetic fluctuations that may persist even at temperatures well above the magnetic ordering temperature and the coupling of defect or surface states. Due to chemical pressure and increased bonding interaction, the resistivity of Eu\textsubscript{5}In\textsubscript{4/3}Ga\textsubscript{2/3}Sb\textsubscript{6} single crystals is orders of magnitude lower than the parent Eu\textsubscript{5}In\textsubscript{2}Sb\textsubscript{6}. In the parent Eu\textsubscript{5}In\textsubscript{2}Sb\textsubscript{6}, we also do not see the \textit{T}$ \sim$60 K feature in the resistivity.\cite{rosa2020colossal} The band gap conceptualized via the Arrhenius model shows \textit{E}\textsubscript{g}=52 meV. Overall, the less resistive behavior is indicative of the contraction of the In/Ga-Sb bond and the distortion in the polyanionic polyhedral of [In\textsubscript{2}Sb\textsubscript{6}]\textsuperscript{10-} that may imply that Eu\textsubscript{5}In\textsubscript{2-x}Ga\textsubscript{x}Sb\textsubscript{6} with  x=$\frac{2}{3}$ getting close to the topological phase transition. 
\begin{figure}
    \centering
    \includegraphics[width=0.3\linewidth]{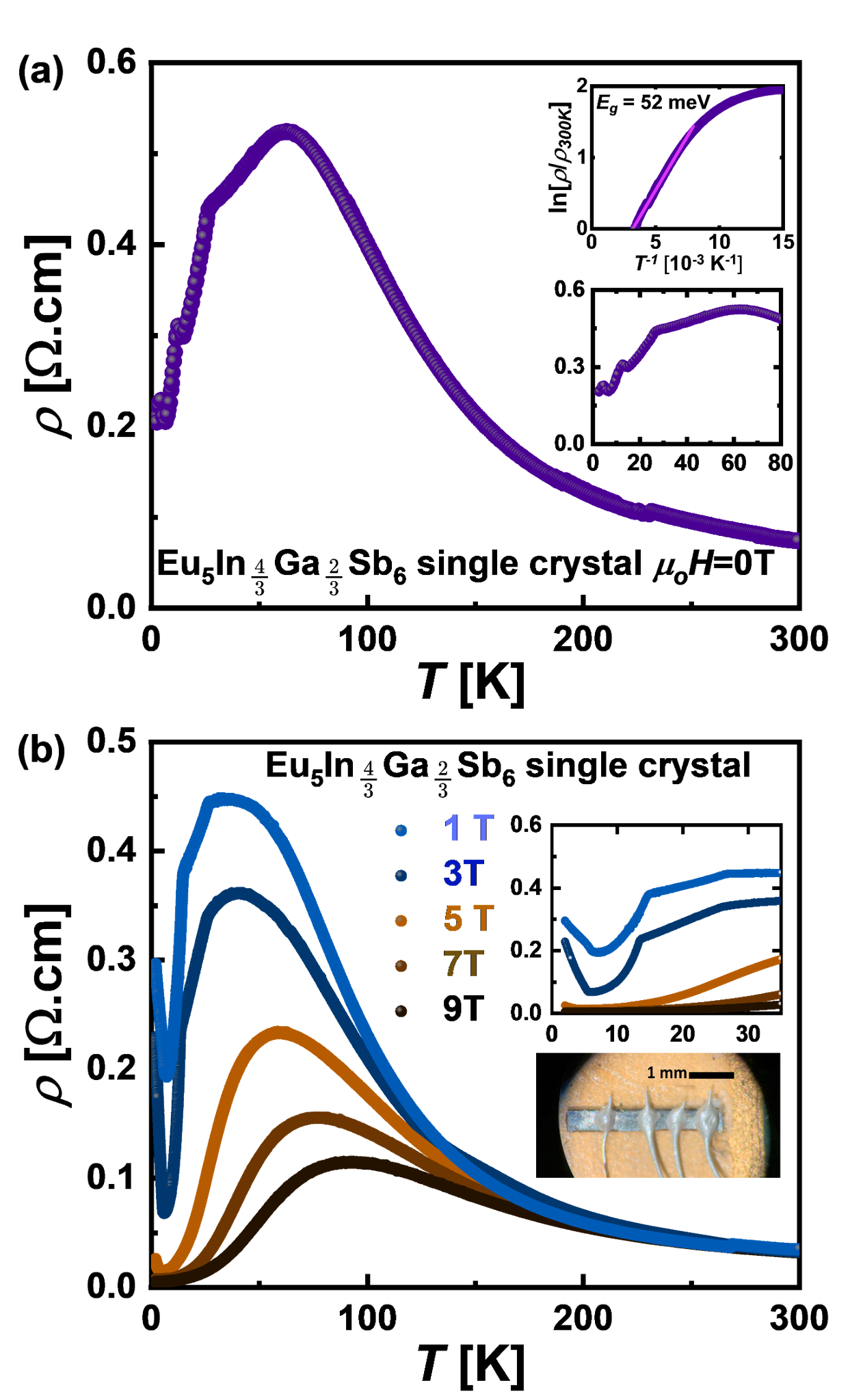}
    \caption{(a) Electrical resistivity of x=$\frac{2}{3}$ Eu\textsubscript{5}In\textsubscript{2-x}Ga\textsubscript{x}Sb\textsubscript{6} single crystal ( i.e. sample-3 from Figure-4) at 0.1 T applied field at the c-axis for applied current. The first inset describes the magnetic transitions due to magnetic order as previously seen in heat capacity and magnetization plots. The second inset describes the gap of 52 meV which is lower than Eu\textsubscript{5}In\textsubscript{2}Sb\textsubscript{6}. (b) Resistivity at a few different applied magnetic fields and the shift conductivity as the applied magnetic field increases. }
    \label{rho}
\end{figure}

\begin{figure}
    \centering
    \includegraphics[width=1\linewidth]{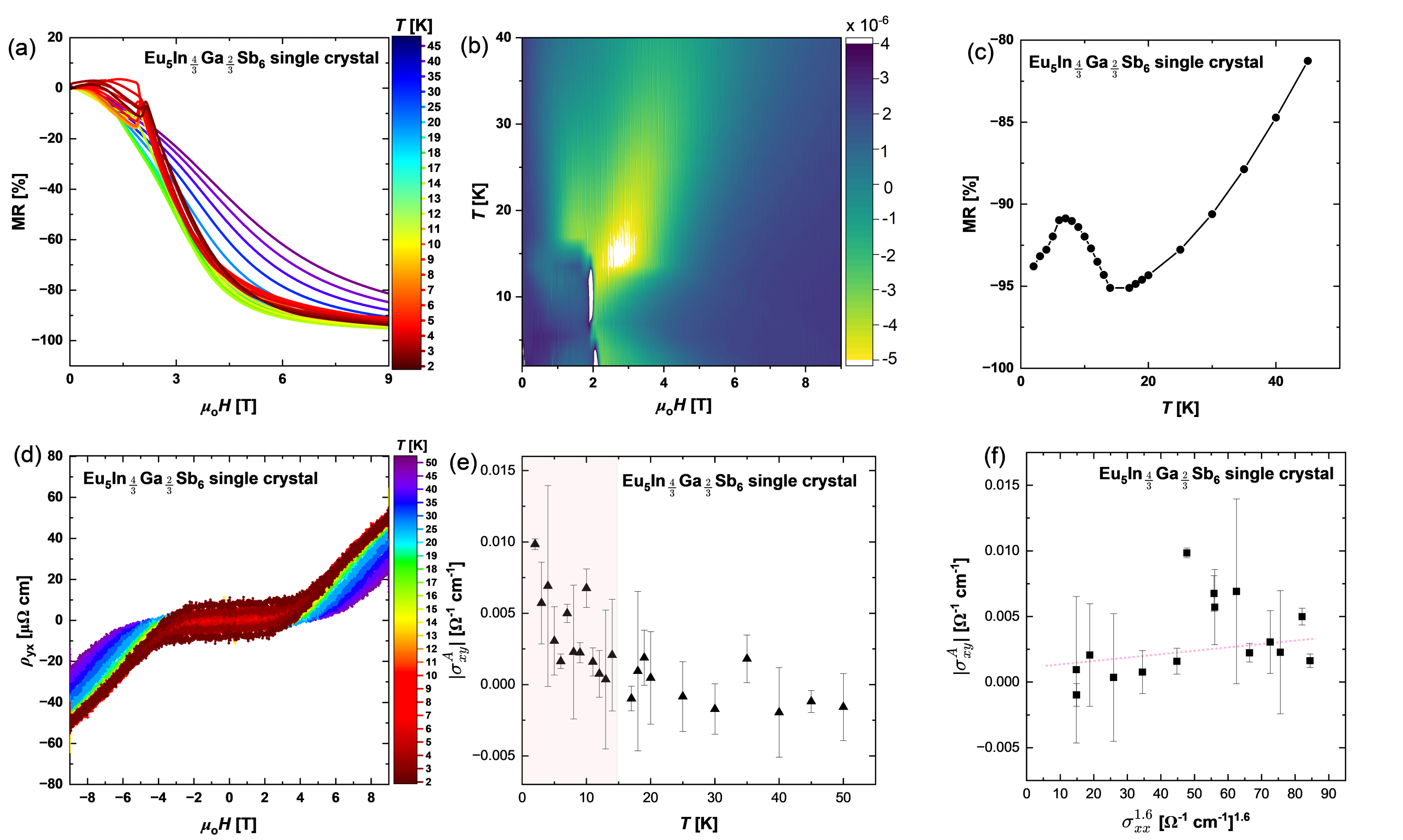}
    \caption{(a) Isothermal magnetoresistance of Eu\textsubscript{5}In\textsubscript{4/3}Ga\textsubscript{2/3}Sb\textsubscript{6} single crystals measured (i.e. sample-3 from Figure-4) for temperatures 2K$<${$T$}$<$45K  and magnetic fields 0T$<$$\mu_{o}\textit{H}$$<$9T for $\mu_{o}\textit{H}$$\perp$b and \textit{j}//b. (b) Contour plot of the field-dependent derivative, (c) temperature dependence of the MR at $\mu_{o}\textit{H}$=9 T, (d) Antisymmetrized Hall resistivity $\rho$$_y$$_x$ data measured corresponding to panel (a). (e) Shows onset of \textit{T}\textsubscript{N}, where \textit{T}\textsubscript{N1}= 15.0 K, \textit{T}\textsubscript{N2}= 5.7 K at $\mu_{o}\textit{H}$=0.1 T. (f) Clarifies the dependence of $ \sigma^A_{xy}$ on  $ \sigma^{1.6}_{xy}$ for \textit{T}$<$\textit{T}\textsubscript{N} respectively. }
    \label{MR}
\end{figure}

To further corroborate the temperature-dependent transport behavior, the Hall resistivity and magnetoresistance were measured. In Figure 6 (a) we see the isothermal magnetoresistance of Eu\textsubscript{5}In\textsubscript{4/3}Ga\textsubscript{2/3}Sb\textsubscript{6} single crystals measured for temperatures 2K$<${$T$}$<$45K  and magnetic fields 0T$<$$\mu_{o}\textit{H}$$<$9T for $\mu_{o}\textit{H}$$\perp$b and \textit{j}//b. The magnetoresistance (MR) is highly sensitive to metamagnetism, which is highlighted in the contour plot of the field-dependent derivative shown in Figure 6 (b). Now, focusing on the high-field data, Figure 6 (c) shows the temperature dependence of the MR at $\mu_{o}\textit{H}$=9 T. Similar to the parent compound, Eu\textsubscript{5}In\textsubscript{2}Sb\textsubscript{6}, a colossal negative MR is registered, ranging from MR($\mu_{o}\textit{H}$=9 T) $\sim$ -80\% at $T$=45 K before monotonically increasing on cooling to a maximum of MR($\mu_{o}\textit{H}$=9 T) $\sim$-94\% at $T$=17 K, just above \textit{T}\textsubscript{N}. Upon further cooling below \textit{T}\textsubscript{N}, MR(9 T) decreases to $\sim$-90\% at \textit{T}\textsubscript{N2} before, again increasing to $\sim$-94 at \textit{T}=2 K. In Eu\textsubscript{5}In\textsubscript{2}Sb\textsubscript{6}, the colossal negative MR was speculated to originate from the formation of magnetic polarons.\cite{rosa2020colossal} 
Our Hall data, presented in Figure 6(d) provides further evidence of possible magnetic polarons below \textit{T}\textsubscript{N}. 

Figure 6(d) further shows the corresponding Hall resistivity $\rho$$_y$$_x$ data measured and antisymmerterized for temperatures 2K$<$${T}$$<$45K and magnetic fields -9T$<$$\mu_{o}\textit{H}$$<$9T for \textit{$\mu$}\textsubscript{o}\textit{H} perpendicular to b and \textit{j}//b. For \textit{T}$>$\textit{T}\textsubscript{N}, $\rho$$_y$$_x$ shows a clear multi-band character with dominant hole-like conduction, evidenced by the positive (negative) curvature and sign in positive (negative) measured fields.

Upon cooling below \textit{T}\textsubscript{N} the $\rho$$_y$$_x$ has a similar behavior at high fields, however, a spontaneous anomalous Hall effect is measured which is observed as $\rho$$_y$$_x$($\mu_{o}\textit{H}$=0T)$\neq0$ in\textbf{ }Figure 6(d). The temperature dependence of the anomalous Hall conductivity $\sigma^A_{xy}= 
  \frac{-\rho_{yx}(\mu_oH=0T)}{\rho^2_{xx}+{\rho^2_{yx}}}
$ is shown in Figure 6(e), clearly demonstrating the onset below \textit{T}\textsubscript{N}, where \textit{T}\textsubscript{N1}= 15.0 K, \textit{T}\textsubscript{N2}= 5.7 K at $\mu_{o}\textit{H}$=0.1 T.
To clarify the origin of the anomalous Hall, we plot the dependence of $ \sigma^A_{xy}$ on  $ \sigma^{1.6}_{xy}$ for \textit{T}$<$\textit{T}\textsubscript{N} in Figure 6(f). In the hopping regime corresponding to  $\sigma_{xx}= 
  \frac{\rho_{xx}}{\rho^2_{xx}+{\rho^2_{yx}}}<10^4  (\Omega.cm)^{-1}
$, such a power law behavior is expected due to the percolation of magnetic clusters.\cite{nagaosa2010anomalous}\cite{liu2011scaling}
By contrast, in the metallic regime where Berry curvature is expected to dominate,  $ \sigma^A_{xy}$ is expected to be independent of impurity scattering, that is  $ \sigma^A_{xy}$ = $constant$. In the most metallic systems $( \sigma_{xy}>10^6 (\Omega.cm)^{-1})$, skew-scattering dominates and  $ \sigma^A_{xy}$$\sim$ $\sigma_{xy}$.\cite{nagaosa2010anomalous} 
Though the data is noisy, Figure 6(f) shows that $ \sigma^A_{xy}$, is clearly not independent of $ \sigma^A_{xx}$, and the exponent is certainly greater than 1, consistent with the percolation of magnetic clusters, like magnetic polarons. However, the hysteresis observed in  $ \rho_{xy}$, Figure-6(c), is not noticed in \textit{M}(\textit{$\mu$}\textsubscript{0}\textit{H}) which suggests that there may be an intrinsic mechanism that derives the anomalous Hall effect. Still, a Berry-phase-induced anomalous Hall, which would be a sign of topology, is not seen, consistent with the theoretical prediction for this substitution level.

\section{Conclusion}
We explore the effect of chemical pressure to induce a topological phase transition in the previously reported topologically trivial magnetic insulator Eu\textsubscript{5}In\textsubscript{2}Sb\textsubscript{6} with $Z$$_2$=0. Although Eu\textsubscript{5}Ga\textsubscript{2}Sb\textsubscript{6} had been predicted to exist with a $Z$$_2$=1, it is synthetically challenging to make and structurally difficult to stabilize in the \textit{Pbam} structure. We thus investigated the effect of Ga substitution in Eu\textsubscript{5}In\textsubscript{2-x}Ga\textsubscript{x}Sb\textsubscript{6} with x=0, $\frac{2}{3}$, $\frac{4}{3}$, and 2 to see where the topological phase transition occurs between the two end compositions Eu\textsubscript{5}In\textsubscript{2}Sb\textsubscript{6} with $Z$$_2$=0 and Eu\textsubscript{5}Ga\textsubscript{2}Sb\textsubscript{6} with $Z$$_2$=1. Per our DFT results utilizing PBE+SOC, we saw that there is a change in parity that occurs between x=$\frac{2}{3}$ and x=$\frac{4}{3}$ (or between x=$\frac{4}{3}$ and x=2 when using the mBJ functional). Crystal Orbital Hamilton Population calculations show that there is a mixture of cationic and polyanionic framework that contributes to the band inversion via changes in Eu-Sb bonding driven by In-Sb/Ga-Sb substitution. We thus, attribute the changes to the third-party or complex hybridization-mediated band inversion effect. Our work introduces a possible magnetic topological insulator via band engineering and explains why Eu-based Zintl compounds are suitable for the co-existence of magnetism and topology.

\section{Experimental Section}
\subsection{Single Crystal Growth, X-ray Diffraction, and X-ray Energy Dispersive Spectroscopy}
The single crystals of  x=$\frac{2}{3}$ Eu\textsubscript{5}In\textsubscript{2-x}Ga\textsubscript{x}Sb\textsubscript{6} series were grown via InSb binary flux and were air-stable on a benchtop for months. For sample-3 with the composition of  x=$\frac{2}{3}$ Eu\textsubscript{5}In\textsubscript{2-x}Ga\textsubscript{x}Sb\textsubscript{6}, the Eu:In:Ga:Sb ratios were 1.0550, 0.8801, 0.57355, and 2.5411 respectively, with a total mass of  
5.0498 grams. The elements Eu (ingot, Yeemeida Technology Co., LTD 99.995\%), In (shots, Sigma-Aldrich 99.99\%), Ga (ingot, Noah Tech 99.99\%), and Sb (BTC, 99.999\%) were utilized for the synthesis. In, Ga, and Sb were loaded in a Canfield crucible (size: 2 mL) at atmospheric conditions, while Eu was loaded in the crucible in an Ar-filled glove box. The Canfield crucible was placed in a quartz ampoule with quartz wool below and above the crucible, evacuated, and sealed under 5.4×10\textsuperscript{–2} Torr of pressure. The evacuated ampoules were loaded in a box furnace at an angle of $\sim$45° with the charge facing the center on the backside of the central interior of the box furnace. The temperature was ramped at a rate of 80°C/hour to 550°C and dwell for 12 hours which allowed for the Ga and InSb fluxes to be in the liquid state. Then the box furnace temperature was ramped at the rate of 80°C/hour to 1100°C and dwelled for 24 hours to allow the Eu\textsubscript{5}In\textsubscript{2-x}Ga\textsubscript{x}Sb\textsubscript{6} to dissolve in the fluxes. The furnace was then cooled at the rate of 5°C/hour to 650°C for the dissolved phase to crystallize on cooling and then centrifugation separate the flux mixtures from the single crystals. The single crystals had a flat rod-like morphology, 1.5-2 mm in width and 4-12 mm in length.

Single Crystal X-ray diffraction and X-ray energy dispersive spectroscopy (EDS) were used to confirm the phase purity and elemental composition of the single crystals. Single crystal X-ray diffraction data were collected using a SuperNova diffractometer equipped with an Atlas detector and a Mo K$\alpha$ source. The cuboid crystal, cut from a larger crystal piece, was mounted with Paratone-N oil. Data was analyzed and reduced using the CrysAlisPro software suite, version 1.171.36.32 (2013), Agilent Technologies. Initial structural models were developed using SIR92 and refinements of this model were done using SHELXL-97 (WinGX version, release 97-2).\textsuperscript{2,3} Real-time back reflection Laue X-ray diffraction was used to orient and align the crystals for measurement. Energy-dispersive x-ray spectroscopy (EDX) in a Quanta environmental scanning electron microscope (SEM) equipped with an Oxford EDX detector at 25keV. Homogeneity of composition was observed using the EDX mapping technique, with small errors in composition likely arising from excess flux left on the surface of crystals. 

\subsection{Theoretical Calculations:}
DFT calculations were performed using Quantum Espresso 7.0 \cite{giannozzi2009quantum}\cite{giannozzi2017advanced} with the PBE functional and pseudopotentials obtained from QUANTUM ESPRESSO.\cite{giannozzi2020quantum} To accommodate variable occupancies on the In site, the reported $Pbam$ unit cell of Eu\textsubscript{5}In\textsubscript{2}Sb\textsubscript{6} was tripled along the crystallographic c axis to produce a new cell $\sim$~11.6x15.1x12.8 Å that retains $Pbam$ symmetry but has two crystallographically distinct In sites (4h and 8i). All calculations were carried out using this enlarged cell for Eu\textsubscript{30}In\textsubscript{12}Sb\textsubscript{36} (x = 0), Eu\textsubscript{30}In\textsubscript{8}Ga\textsubscript{4}Sb\textsubscript{36} (x=$\frac{2}{3}$), Eu\textsubscript{30}In\textsubscript{4}Ga\textsubscript{8}Sb\textsubscript{36} (x=$\frac{4}{3}$), and Eu\textsubscript{30}Ga\textsubscript{12}Sb\textsubscript{36} (x = 1). Initially, variable cell relaxations were carried out to minimize the forces and stresses in fully relativistic calculations with spin-orbit coupling and non-colinear magnetization with time-reversal symmetry (no magnetic order). Relaxations were performed using a 4x3x4 Monkhorst-Pack mesh, a 55 Ry kinetic energy cutoff, a 440 Ry charge density cutoff, and Marzari-Vanderbilt-DeVita-Payne cold smearing\cite{marzari1999thermal}, and converged to an energy threshold of 7.8·10\textsuperscript{-4}, and force threshold of 10\textsuperscript{-4}. This resulted in the cell parameters given in Table-S1. The relaxed unit cells and atomic positions were then fixed and used for self-consistent-field calculations using a 6x4x6 Monkhorst-Pack mesh, a 55 Ry kinetic energy cutoff, a 440 Ry charge density cutoff, and Marzari-Vanderbilt-DeVita-Payne cold smearing \cite{marzari1999thermal}, and converged to an energy error of 1.5·10\textsuperscript{-8} Ry. These SCF calculations were carried out for both the spin-orbit-coupled and non-spin-orbit-coupled cases. Band unfolding to the original (non-enlarged) cell was carried out using a publicly available tools at GitHub. Symmetry analysis of bands at the time-reversal invariant momenta was carried out using the post-processing tools of Quantum Espresso.\cite{giannozzi2009quantum}\cite{giannozzi2017advanced} Crystal Orbital Hamilton Population (COHP) analysis for the non-SOC calculations was carried out using LOBSTER.\cite{maintz2013analytic} COHP for the SOC case was carried out using the same framework\cite{maintz2016lobster} implemented as a custom Python code and checked for consistency with the non-SOC LOBSTER results. The DFT MBJ+SOC calculations were performed using the converged lattice parameters from PBE+SOC as given in Table-S1. 

\subsection{Bulk Magnetic Properties Measurements}
Magnetization data were collected on a Quantum Design magnetic property measurement system (MPMS). Magnetic susceptibility was approximated as magnetization divided by the applied magnetic field ($\chi$$\approx$$M/H$). Magnetization measurements were performed in a vibrating sample superconducting quantum interference device magnetometer (SQUID-VSM) from Quantum Design. All measurements were carried out after cooling in a zero field. To reduce the remnant field of the superconducting magnet to less than 2 Oe before each measurement, we applied a magnetic field of 7 T at ambient temperature and then removed it in an oscillation mode. The magnetic field was applied to the a, b, and c axis to the flat rod-like crystal. Sample shape correction was accounted for in these measurements. 

\subsection{Thermodynamic Properties}
Heat capacity measurements were performed in a Quantum Design Physical Properties Measurement System (PPMS). Measurements were performed on a single crystalline sample of x=$\frac{2}{3}$ Eu\textsubscript{5}In\textsubscript{2-x}Ga\textsubscript{x}Sb\textsubscript{6} of 2.9(3) mg mass oriented such that the applied magnetic field was along the nominal c-axis to the current. The heat capacity was measured $T$=2-300K and various applied fields. The sample was mounted on the sample stage using Apiezon N grease.

\subsection{Electrical Transport Properties:}
The resistivity option in the PPMS-9 was utilized to carry out the resistivity measurements. The resistivity was measured from \textit{T} = 2-300 K using the four-probe technique at various applied magnetic fields. The leads were made from Pt wire and the contacts were made using Dupont 4922N Ag paste. The Pt lead distance was 0.38 mm. The sample length was 1.3 mm longitudinally.

Electrical-magneto transport measurements were performed in a Quantum Design Dynacool Physical Properties Measurement System using the Electrical Transport Option. Contacts to the sample were made in a standard four-point or Hall bar geometry. Gold wire 0.025m annealed (metal basis) was utilized for the electrodes and Ag conductive paint SPI\# 05002-AB was utilized for the contacts. For magnetoresistance and Hall measurements, full hysteresis loops were measured such that Quadrant I denoted as QI was measured from $\mu_oH>0 T \rightarrow 0 T$, QII from $ 0 T \rightarrow \mu_oH<0 T$, QIII from $ \mu_oH<0 T \rightarrow 0 T$, and QIV from $0 T \rightarrow \mu_oH>0T$.   The data were symmetrized or antisymmetrized, respectively such that 
\begin{eqnarray}{\rho_{xx}(\text{d}H>0,\text{d}H<0}) =\frac{{R_{meas}(QIV,QI)}+{R_{meas}(QII,QIII)}}{2}\frac{wt}{L}\end{eqnarray}
\begin{eqnarray}{\rho_{yx}(\text{d}H>0,\text{d}H<0}) =\frac{{R_{meas}(QIV,QI)}-{R_{meas}(QII,QIII)}}{2}t.\end{eqnarray}
Here, $R_{meas}$ is the resistance of the raw measured data, while \textit{L} is the length between voltage contacts, \textit{w} the width of the sample, and \textit{t} the sample thickness. 

\medskip
\textbf{Supporting Information} The supporting information consists of materials and method section, addition result and discussion, addition DFT results including pCOHP/bond(energy), single crystal refinement tables, supercell relaxed lattice parameters, magnetic properties, thermodynamic properties, and SEM-EDX results in the Eu\textsubscript{5}In\textsubscript{2-x}Ga\textsubscript{x}Sb\textsubscript{6} series.

\medskip
\textbf{Acknowledgements} \par 
T.B. acknowledges postdoctoral support from NSF MRSEC through the Princeton Center for Complex Materials NSF-DMR-2011750. LMS was supported by the Gordon and Betty Moore Foundation’s EPIQS initiative through Grants GBMF9064, The Packard foundation, and the Princeton Catalysis Initiative (PCI). The authors thank Rafal Wawrzyńczak, Chris Lygouras, and Will Liag for their helpful discussions. SBL is supported by the National Science Foundation Graduate Research Fellowship Program under Grant No. DGE-2039656. AZ and DS acknowledge support from NSF SSMC award number no. 1709158\textit{.} TMM acknowledges support from by the Institute for Quantum Matter, an Energy Frontier Research Center funded by the U.S. Department of Energy, Office of Science, Office of Basic Energy Sciences, under Grant DE-SC0019331. The MPMS was funded by the National Science Foundation, Division of Materials Research, Major Research Instrumentation Program, under Award No. 1828490. Any opinions, findings, and conclusions or recommendations expressed in this material are those of the author(s) and do not necessarily reflect the views of the National Science Foundation.

\medskip
\textbf{Conflict of interest} \par
The authors declare no conflict of interest.

\medskip
\textbf{Data availability statement} \par
The data that support the findings of this study are available from the corresponding author upon reasonable request.\\

\medskip


\newpage

\setcounter{figure}{0}
\setcounter{table}{0}
\setcounter{equation}{0}
\setcounter{page}{1}
\setcounter{section}{0}

\centerline{\bfseries \LARGE Supplementary Information}\setlength{\parskip}{12pt}%

\begin{table}
\begin{tabular}{lr}
 &Page Number\\
 Result and Discussion of Magnetic and Thermodynamic Properties&S-3 to S-5\\
 Table S1: Eu\textsubscript{5}In\textsubscript{2-x}Ga\textsubscript{x}Sb\textsubscript{6}: 3c supercell relaxed lattice parameters&S-6\\
 Table S2: The parity outputs at the k points using MBJ in Eu\textsubscript{5}In\textsubscript{2-x}Ga\textsubscript{x}Sb\textsubscript{6}&S-7\\
 Figure S1: DFT plots using GGA+SOC for the Eu\textsubscript{5}In\textsubscript{2-x}Ga\textsubscript{x}Sb\textsubscript{6}&S-8\\
 Figure S2: DFT plots using MBJ for the Eu\textsubscript{5}In\textsubscript{2-x}Ga\textsubscript{x}Sb\textsubscript{6}&S-9\\
 Figure S3: The pCOHP/bond as a function of energy that compares dimers&S-10\\
 Table S3: Crystal data and structure refinement for Eu$_5$In$_{1.37}$Ga$_{0.63}$Sb$_6$&S-11\\
    Table S4: Displacement parameters for Eu$_5$In$_{1.37}$Ga$_{0.63}$Sb$_6$ structure.& S-12\\
    Table S5: Anisotropic displacement parameters for Eu$_5$In$_{1.37}$Ga$_{0.63}$Sb$_6$ & S-13\\
 Figure S4: Magnetism in Eu$_5$In$_{1.37}$Ga$_{0.63}$Sb$_6$  &S-14\\
 Figure S5: Heat Capacity in Eu$_5$In$_{1.37}$Ga$_{0.63}$Sb$_6$  &S-15\\
    Figure S6: Shift structural in Eu$_5$In$_{1.37}$Ga$_{0.63}$Sb$_6$ & S-16\\
 Figure S7: SEM-EDS maps and spectrum of Eu$_5$In$_{1.586}$Ga$_{0.429}$Sb$_6$&S-17\\
 Figure S8: SEM-EDS maps and spectrum of Eu$_5$In$_{1.495}$Ga$_{0.637}$Sb$_6$&S-18\\
 Figure S9: SEM-EDS maps and spectrum of Eu$_5$In$_{1.378}$Ga$_{0.663}$Sb$_6$&S-19\\
 Figure S10: The DFT Supercell positions of x=2 in Eu\textsubscript{5}In\textsubscript{2-x}Ga\textsubscript{x}Sb\textsubscript{6}.&S-20\\
 Figure S11: The DFT Supercell positions of x=4/3 in Eu\textsubscript{5}In\textsubscript{2-x}Ga\textsubscript{x}Sb\textsubscript{6}.&S-21\\
 Figure S12: The DFT Supercell positions of x=4/3 in Eu\textsubscript{5}In\textsubscript{2-x}Ga\textsubscript{x}Sb\textsubscript{6}.&S-22\\
 Figure S13: The DFT Supercell positions of x=0 in Eu\textsubscript{5}In\textsubscript{2-x}Ga\textsubscript{x}Sb\textsubscript{6}.&S-23\\
 Figure S14: The DFT Supercell structures for x=0, $\frac{2}{3}$, $\frac{4}{3}$, and 2 in Eu\textsubscript{5}In\textsubscript{2-x}Ga\textsubscript{x}Sb\textsubscript{6}.&S-24\\
 References&S-25\\
  \end{tabular}
\end{table}
\pagebreak
\section{Supplementary Results and Discussion}
\subsection{Magnetic Properties}
To examine the direction-dependent magnetic anisotropy of Eu\textsubscript{5}In\textsubscript{2-x}Ga\textsubscript{x}Sb\textsubscript{6} with  x=$\frac{2}{3}$, the magnetization as a function of field and temperature was studied. In concurrence with the heat capacity measurement, the \textit{T}\textsubscript{N1} and \textit{T}\textsubscript{N2} are the same i.e. \textit{T}\textsubscript{N1}= 15.029 K, \textit{T}\textsubscript{N2}= 5.720 K at $\mu\textsubscript{o}\textit{H}$=0.1 T in $\mu\textsubscript{o}\textit{H}$//a, b, and c directions. In addition to the two transitions, there is another weak  transition that occurs below \textit{T}\textsubscript{N2}, i.e. \textit{T}\textsubscript{N3}= 3.483 K.

Notably, there is a difference in the magnitude of magnetization as a function of temperature in each of the three directions, as seen in Figure S4 (a)-(c). The b axis has the lowest magnetization compared to the a and c axis. Metamagnetism in Figure S4 (d)-(e) indicates AFM order polarized within each ab plane. The c-axis is the hard magnetic axis with no component of the AFM order parameter polarized along the c-plane. Therefore, from the perspective of local spins the b-axis is considered the easy axis in antiferromagnets and hard axis in ferromagnets.\cite{yosida1996theory} The non-saturating moment of Eu\textsuperscript{2+} state is indicative of strong \textit{J} coupling between the Eu\textsuperscript{2+} cations and the requirement of stronger applied field than 7 T to saturate the Eu\textsuperscript{2+} magnetic moment. 

The metamagnetism in $\mu\textsubscript{o}\textit{H}$//a and b is due to a spin flip transition that occurs in the ab plane. In general metamagnetism is seen in most Gd\textsuperscript{3+} and Eu\textsuperscript{2+} compounds because of high single-ion isotropy due to the isotropic nature of a 4\textit{f}\textsuperscript{7} (\textit{S} = 7/2) state.\cite{holmes1966magnetic} In Eu\textsubscript{5}In\textsubscript{2-x}Ga\textsubscript{x}Sb\textsubscript{6} with  x=$\frac{2}{3}$, the magnetic anisotropy dominates the polarizability of the spin in the a and b direction compared to the c direction, therefore in agreement with our previous claim on c axis being the hard axis. 

Overlaying the transitions coming from \textit{M}(\textit{T}) and \textit{M}($\mu\textsubscript{o}\textit{H}$) in a, b, and c directions, we can construct the magnetic phase diagrams in Figure S4 (g-i). We gather that the overall magnetism in Eu\textsubscript{5}In\textsubscript{2-x}Ga\textsubscript{x}Sb\textsubscript{6} with  x=$\frac{2}{3}$ is antiferromagnetic and the spins lie in the ab plane. Thus, there are four distinct magnetic phases in the a and b axis, where two magnetic phases in the c direction respectively. The two magnetic phases in the c direction correspond to the two magnetic phases corresponding to \textit{T}\textsubscript{N1} and \textit{T}\textsubscript{N2}. These phases per the single crystal neutron diffraction correspond to the \textit{k}=(0,0,0) and \textit{k}=(0,0,½), where at \textit{T}\textsubscript{N1} the unit cell expands.\cite{morano2024noncollinear2} The overall magnetic structure is proposed to be similar to the parent Eu\textsubscript{5}In\textsubscript{2-x}Ga\textsubscript{x}Sb\textsubscript{6} with non-collinear magnetism at \textit{T}=2K, since the Eu\textsubscript{5}In\textsubscript{2-x}Ga\textsubscript{x}Sb\textsubscript{6} with  x=$\frac{2}{3}$  has the same qualitative features in magnetism. \cite{morano2024noncollinear2}
\subsection{Thermodynamic Properties}
The heat capacity of in Eu\textsubscript{5}In\textsubscript{2-x}Ga\textsubscript{x}Sb\textsubscript{6} with  x=$\frac{2}{3}$  shows two transitions at from \textit{T}\textsubscript{N1}= 5.529(6) K, \textit{T}\textsubscript{N2}=14.415(2)K as seen in Figure S5(a). Previous work on the parent Eu\textsubscript{5}In\textsubscript{2}Sb\textsubscript{6} structure attributed these transitions to be second-order phase transitions due to the antiferromagnetic order of Eu.\cite{rosa2020colossal2} In  Eu\textsubscript{5}In\textsubscript{2-x}Ga\textsubscript{x}Sb\textsubscript{6} with x=$\frac{2}{3}$, these transitions are qualitatively similar to the parent. \textit{T}\textsubscript{N1} gets suppressed on applying a magnetic field of $\mu\textsubscript{o}\textit{H}$=9T, however, the \textit{T}\textsubscript{N2} requires a 
$\mu\textsubscript{o}\textit{H}>$ 9 T to suppress the magnetic order in Eu\textsubscript{5}In\textsubscript{2-x}Ga\textsubscript{x}Sb\textsubscript{6} with  x=$\frac{2}{3}$ as seen in Figure S5(c).

In the parent version, the magnetic order for \textit{T}\textsubscript{N2} is in agreement with the magnetization as a function of field behavior in Eu\textsubscript{5}In\textsubscript{2}Sb\textsubscript{6}, where a $\mu\textsubscript{o}\textit{H}$=30T saturates the Eu\textsuperscript{2+} \textit{S}=\textit{J}=7/2 in the c direction.\cite{rosa2020colossal2} The $\Delta$\textit{T}\textsubscript{N2} in the two cases are similar, suggesting that $\mu\textsubscript{o}\textit{H}$→30 T may be expected to saturate Eu\textsuperscript{2+} spins. Requiring a large expected saturation magnetic field in Eu\textsuperscript{2+} spins is indicative of strong  \textit{J} coupling. Typically mostly Eu\textsuperscript{2+} Zintl antiferromagnetic compounds have far weaker \textit{J} coupling i.e. a smaller $\mu\textsubscript{o}\textit{H}$ is sufficient to saturate the magnetic order. \cite{ZintlMat4berry2022type} However, recent neutron studies of Eu\textsubscript{5}In\textsubscript{2}Sb\textsubscript{6} reveal that at \textit{T}=2K the Eu\textsuperscript{2+} spins become non-collinear giving rise to magnetic complexities and contribute to stronger \textit{J} coupling within the Eu sublattice.\cite{morano2024noncollinear2} However, recent neutron studies of Eu\textsubscript{5}In\textsubscript{2}Sb\textsubscript{6} reveal that at \textit{T}=2K the Eu\textsuperscript{2+} spins become non-collinear giving rise to magnetic complexities and contributing to stronger \textit{J} coupling within the Eu sublattice. There are different Eu sublattices in Eu\textsubscript{5}In\textsubscript{2}Sb\textsubscript{6} and have been reported to have a non-collinear magnetic structure via neutron studies. In Eu\textsubscript{5}In\textsubscript{2-x}Ga\textsubscript{x}Sb\textsubscript{6} with  x=$\frac{2}{3}$, we observe a contraction of bond distances between Ga and Eu as seen in Figure 3. There could be a possibility that the more conductive network partly due to Eu-Ga bonds along with the non-collinear structure could contribute to the topology in the more Ga-rich compositions.

To confirm the \textit{S}=\textit{J}=7/2 Eu\textsuperscript{2+} spin state in Eu\textsubscript{5}In\textsubscript{2-x}Ga\textsubscript{x}Sb\textsubscript{6} with  x=$\frac{2}{3}$, an analytical model for the phonon-specific heat capacity in $\frac{2}{3}$Ca\textsubscript{5}Ga\textsubscript{2}Sb\textsubscript{6} + $\frac{1}{3}$Ca\textsubscript{5}In\textsubscript{2}Sb\textsubscript{6} was constructed and the mass difference between Eu and Ca were adjusted using the equation below.\textbf{
\begin{eqnarray}{\frac{\Theta^3_{L_{m}Y_{s}O_{t}Z_{r}}}{{\Theta^3_{W_{m}X_{n}Y_{o}Z_{p}}} }}=\frac{mM^\frac32_L+sM^\frac32_Y+tM^\frac32_O+rM^\frac32_Z}{mM^\frac32_W+nM^\frac32_X+oM^\frac32_Y+pM^\frac32_Z}\end{eqnarray}}

where $\Theta^3_{L_{m}Y_{s}O_{t}Z_{r}}$ and $\Theta^3_{W_{m}X_{n}Y_{o}Z_{p}}$ are the normalization factor of x=2/3 Eu\textsubscript{5}In\textsubscript{2-x}Ga\textsubscript{x}Sb\textsubscript{6} and ($\frac{2}{3}$Ca\textsubscript{5}Ga\textsubscript{2}Sb\textsubscript{6} +$\frac{1}{3}$Ca\textsubscript{5}In\textsubscript{2}Sb\textsubscript{6}), and  are the molar masses of Eu\textsubscript{5}In\textsubscript{1.32}Ga\textsubscript{0.67}Sb\textsubscript{6} and ($\frac{2}{3}$Ca\textsubscript{5}Ga\textsubscript{2}Sb\textsubscript{6} + $\frac{1}{3}$Ca\textsubscript{5}In\textsubscript{2}Sb\textsubscript{6}) as 5Eu+1.32In+0.67Ga+6Sb and [$\frac{2}{3}$(5Ca+2In+6Sb)+ $\frac{1}{3}$(5Ca+2Ga+6Sb)]. The normalization factor outputted is 1.155 respectively.\cite{hofmann1956analysis},\cite{tari2003specific}

The non-magnetic analogs, Ca\textsubscript{5}In\textsubscript{2}Sb\textsubscript{6} and Ca\textsubscript{5}In\textsubscript{2}Sb\textsubscript{6}, utilized for the phonon subtraction in the Eu\textsubscript{5}In\textsubscript{2-x}Ga\textsubscript{x}Sb\textsubscript{6} with  x=$\frac{2}{3}$ were in the polycrystalline form. In Figure S5(b) we see that the \textit{C}\textsubscript{P}/\textit{T} plot where the constructed model of Ca analogs have the same heat capacity at \textit{T}$\sim$125 K suggesting that the Eu\textsubscript{5}In\textsubscript{2-x}Ga\textsubscript{x}Sb\textsubscript{6} with  x=$\frac{2}{3}$ is magnetic beyond its magnetic ordering temperature. At \textit{T}$\ge$125 K the two heat capacities overlap and follow the Dulong Petit law. On subtracting the phonons using the Ca analog, the entropy recovered between \textit{T}=2-300 K saturates close to the 5\textit{ R}ln(8)=86 J.mol\textsuperscript{-1}K\textsuperscript{-2} expected for ordering of Eu\textsuperscript{2+} (\textit{S}= 7/2) ions. The complete saturation of \textit{S}=\textit{J}=7/2 spin for Eu\textsuperscript{2+} spin state confirms no presence of mix-valency due to the presence of Eu\textsuperscript{3+}as well as no splitting in the crystal field levels. \cite{fulde1985magnetic}
\pagebreak
\pagebreak
\begin{table}[H]
  \caption{Eu\textsubscript{5}In\textsubscript{2-x}Ga\textsubscript{x}Sb\textsubscript{6}: 3c supercell relaxed lattice parameters.}
  \label{DFTlp}
  \begin{tabular}{cccc}
    \hline
          x ratios&a (Å)
&b (Å)&  c (Å)\\
    \hline
          x=0&11.6326&15.0771&  13.0537 (4.3512)
\\
          x=$\frac23$&11.6059&15.0158&  12.9895 (4.3298)
\\
          x=$\frac43$&11.587&14.9567&  12.9058 (4.3019)
\\
          x=2&11.574&14.9064&  12.8118 (4.2706)

\\\hline
  \end{tabular}
\end{table}

\pagebreak
\begin{table}[H]
  \caption{The parity outputs at the k points using MBJ+SOC in Eu\textsubscript{5}In\textsubscript{2-x}Ga\textsubscript{x}Sb\textsubscript{6}.}
  \label{parity2}
  \begin{tabular}{ccccccccc}
    \hline
          x ratios&$\Gamma$&          $Y$&$X$&$Z$&$U$&$T$&$S$&  $R$\\
    \hline
          x=0&1&     1&1&1&1&1&1&  1\\
          x=$\frac23$&-1&     1&1&-1&-1&1&1&  -1\\
          x=$\frac43$&1&     1&1&1&1&1&1&  1\\
          x=2&1&     -1&1&-1&-1&-1&1&  -1\\\hline
  \end{tabular}
\end{table}
\pagebreak
\begin{figure}
\includegraphics[scale=0.55]{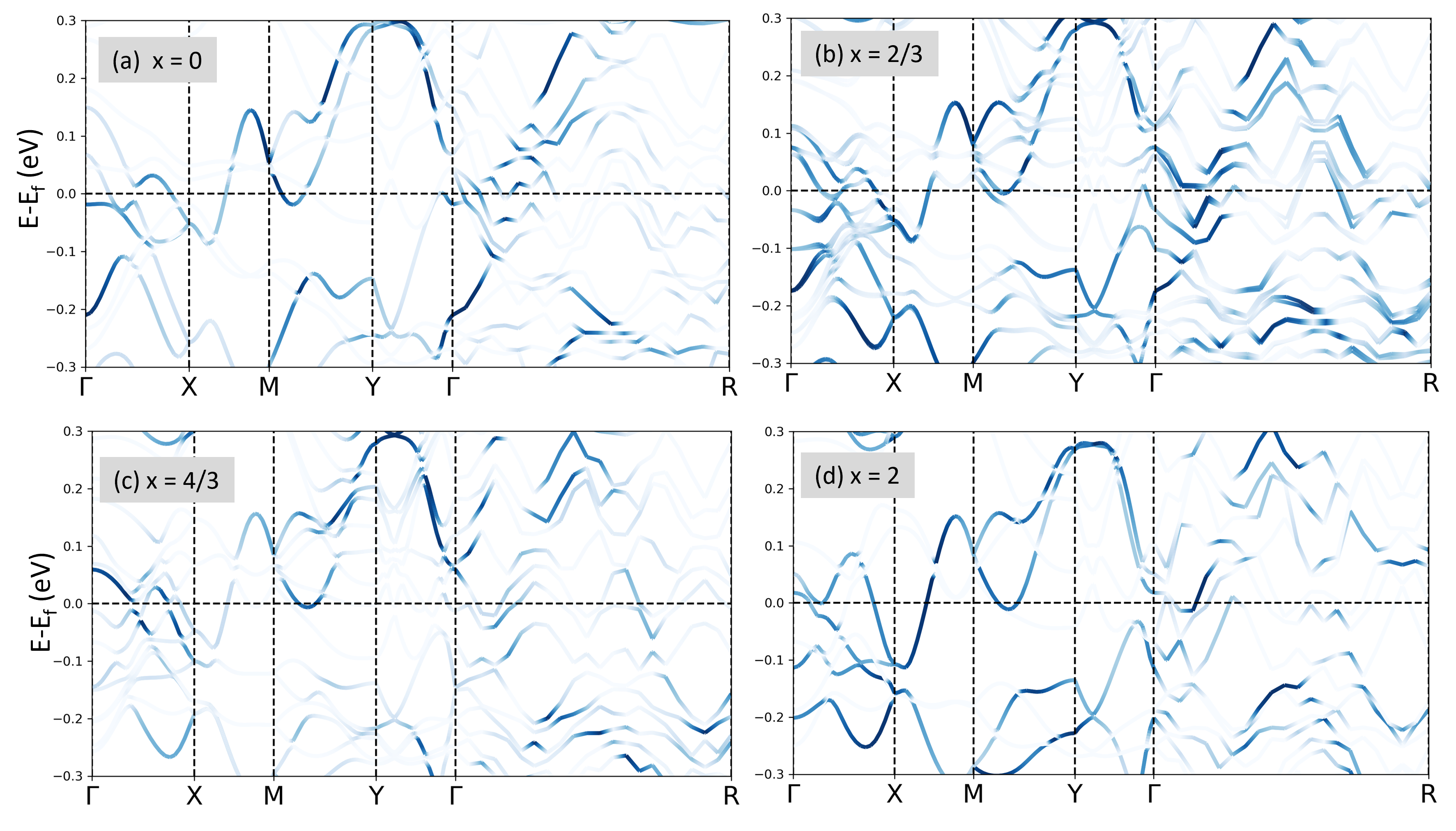}
\caption{Density Functional Theory (DFT) plots using GGA+SOC for the Eu\textsubscript{5}In\textsubscript{2-x}Ga\textsubscript{x}Sb\textsubscript{6} structure for (a) x=0, (b) x=$\frac{2}{3}$, (c) x=$\frac{4}{3}$, and (d) x=2 respectively. The intensity of the dots indicates the fractional contribution of that band to the average structure. }\end{figure}

\pagebreak

\pagebreak
\begin{figure}
\includegraphics[scale=0.78]{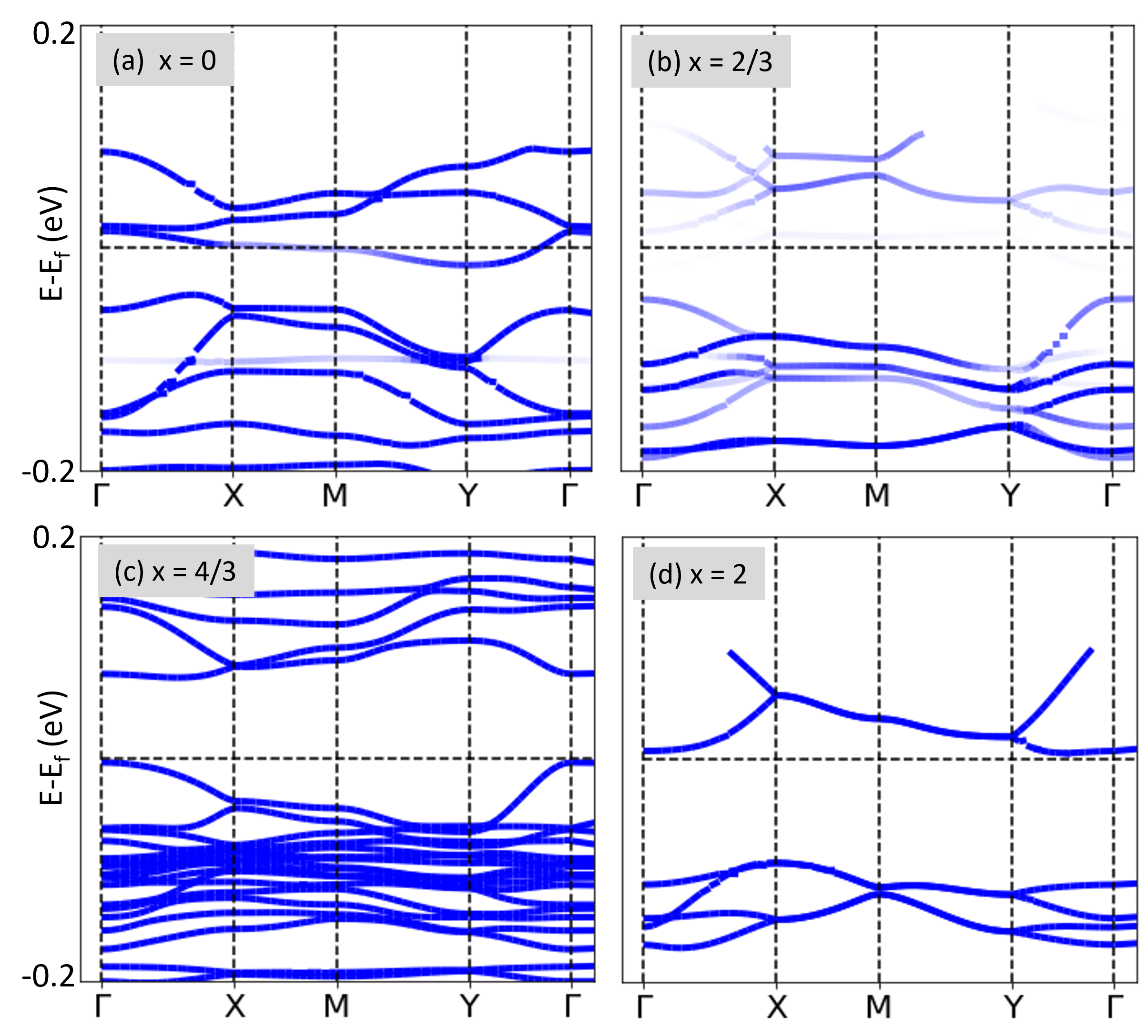}
\caption{\label{Fig-S1.png} Density Functional Theory (DFT) plots using MBJ for the Eu\textsubscript{5}In\textsubscript{2-x}Ga\textsubscript{x}Sb\textsubscript{6} structure for (a) x=0, (b) x=$\frac{2}{3}$, (c) x=$\frac{4}{3}$, and (d) x=2 respectively. The intensity of the dots indicates the fractional contribution of that band to the average structure. }
\label{DFTbig}
\end{figure}
\pagebreak\pagebreak
\begin{figure}
\includegraphics[scale=0.78]{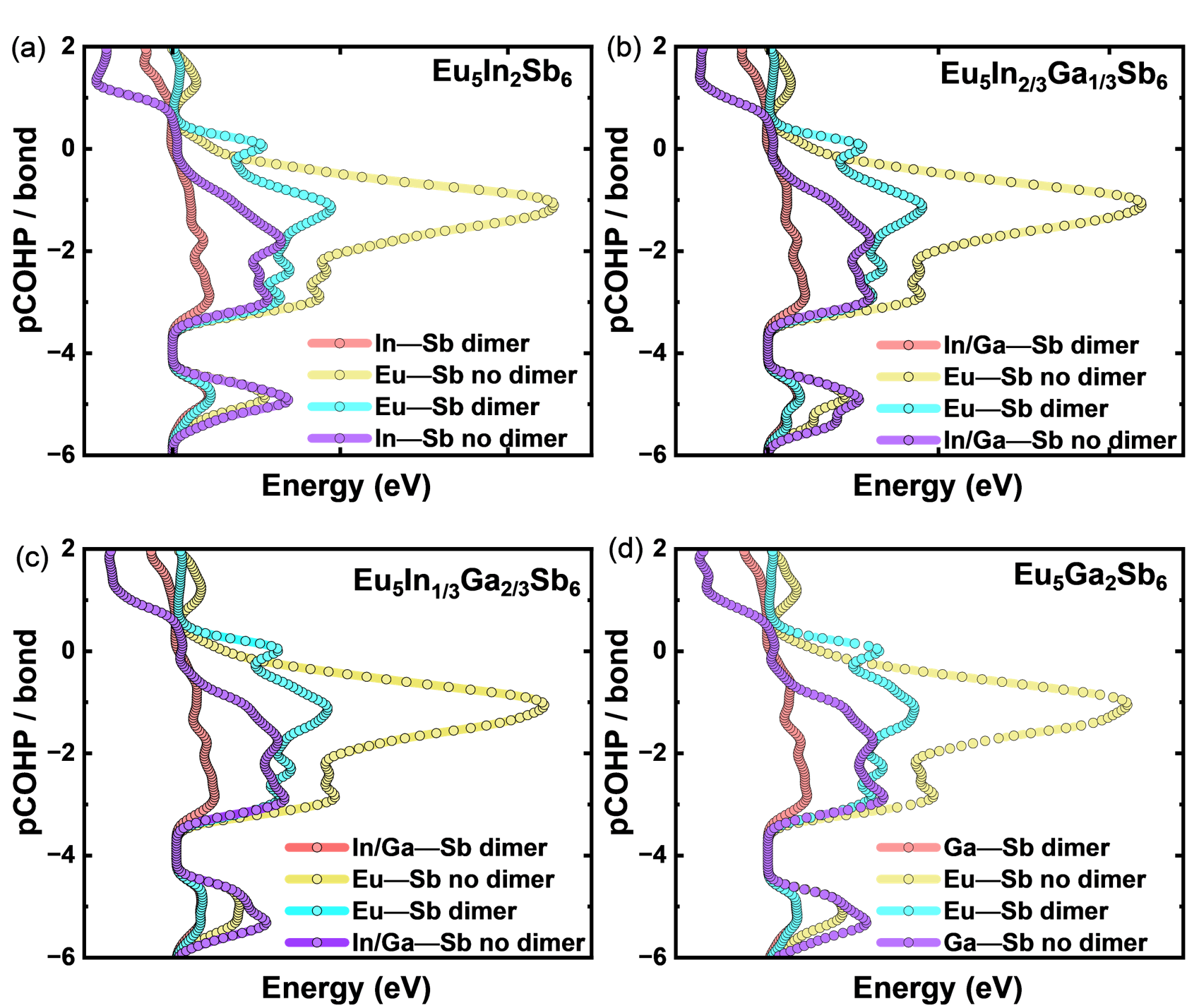}
\caption{\label{pCOHP} The pCOHP/bond as a function of energy that compares the In-Sb dimer, Eu-Sb no dimer, Eu-Sb dimer, and In-Sb no dimer in (a) Eu\textsubscript{5}In\textsubscript{2}Sb\textsubscript6, (b) Eu\textsubscript{5}In\textsubscript{2/3}Ga\textsubscript{1/3}Sb\textsubscript6, (c) Eu\textsubscript{5}In\textsubscript{1/3}Ga\textsubscript{1/3}Sb\textsubscript6, and (d) Eu\textsubscript{5}Ga\textsubscript{2}Sb\textsubscript6. $x=$$\frac{2}{3}$ Eu\textsubscript{5}In\textsubscript{2-x}Ga\textsubscript{x}Sb\textsubscript6 respectively. }\end{figure}

\pagebreak

\pagebreak

\begin{table}[H]
  \caption{Crystal data and structure refinement for Eu$_{5}$In$_{1.37}$Ga$_{0.63}$Sb$_{6}$.}
  \label{SCXRD}
  \begin{tabular}{c c}
    \hline
    Empirical formula &Eu$_{5}$In$_{1.37}$Ga$_{0.63}$Sb$_{6}$\\
    \hline
    Crystal system &Orthorhombic\\
    Space group &\textit{Pbam} (No. 55)\\
    Formula weight &1697.5\\
    a (Å)&14.5200(5)\\
    b (Å)&12.5140(4)\\
    c (Å)&4.5996 (1)\\
    Volume (Å$ ^3$)&835.76 (0)\\
    Z &2 \\
    Temperature (K) &163 (2)\\
    Absorption coefficient (mm$^{-1}$) &30.742\\
    Mo K$\alpha$ (Å)&0.71073\\
    Reflections collected / Number of parameters&1363/45\\
    Goodness of fit&1.082\\
    R[F]$^a$&0.0263\\
    R$_w$(F$ ^2_o$)$ ^b$&0.0345\\
    \hline
    $ ^a R=\frac{\Sigma||F_{o}|-|F_{C}||}{\Sigma|F_{o}|}$&\\
    $ ^b$ R$_w$(F$ ^2_o$)$=[\frac{\Sigma[w(|F_{o}|^{2}-|F_{c}|^{2})^{2}]}{\Sigma[w(|F_{o}|^{4})]}]^{1/2}$&\\
 \hline
  \end{tabular}
\end{table}
  \pagebreak
\begin{table}[H]
  \caption{Fractional atomic coordinates and isotropic displacement parameters based on the refined Eu$_{5}$In$_{1.37}$Ga$_{0.63}$Sb$_{6}$ structure.}
  \label{SCXRD2}
  \begin{tabular}{ccccccc}
    \hline
          Element&Wyckoff Positions&x&  y&z&Occupancy&Uiso\\
    \hline
          Eu1&4h&0.51889(3)          &  0.328445(4)   &0.500000  &1&0.011\\
          Eu2&2d&0.50000&  0.50000&0.50000&1&0.010
\\
          Eu3&4h&0.74940(3)&  0.08761(4)&0.50000&1&0.010
\\
          In1&4g&0.71442(5) &  0.32944(6) &1.0000&0.683(11)&0.013
\\
          Ga1&4g&0.71442(5) &  0.32944(6) &1.0000&0.317(11)&0.013
\\
          Sb1&4g&0.59409(4) &  0.15429(5) &1.0000&1&0.009
\\
          Sb2&4g&0.90374(4) &  0.02325(5) &1.0000&1&0.011
\\
          Sb3&4h&0.82415(5) &  0.33538(5) &0.5000&1&0.015
\\
\hline
  \end{tabular}
\end{table}
\pagebreak
\begin{table}[H]
  \caption{Anisotropic displacement parameters for Eu$_{5}$In$_{1.37}$Ga$_{0.63}$Sb$_{6}$ determined by SXRD.}
  \label{SCXRD3}
  \begin{tabular}{ccccccc}
    \hline
          Element&U(1,1)&U(2,2)&  U(3,3)&U(1,2)&U(1,3)&U(2,3)\\
    \hline
          Eu1&0.0135(2) &0.0092(2)&  0.0096(2) &0.000&0.000&0.00246(18)\\
          Eu2&0.0104(3)&0.0081(3)&  0.0120(3)&0&0&-0.0010(2)
\\
          Eu3&0.0088(2)&0.0092(2)&  0.0119(2)&0&0&-0.00050(17)
\\
          In1&0.0115(3)&0.0094(3)&  0.0073(3)&0&0&-0.0016(2)
\\
          Ga1&0.0088(3)&0.0126(3)&  0.0106(3)&0&0&0.0009(2)
\\
          Sb1&0.0127(3)&0.0075(3)&  0.0244(4)&0&0&-0.0013(2)
\\
          Sb2&0.0126(4)&0.0121(4)&  0.0128(4)&0&0&-0.0003(3)
\\
          Sb3&0.0126(4)&0.0121(4)&  0.0128(4)&0&0&-0.0003(3)
\\
\hline
  \end{tabular}
\end{table}
\pagebreak
\begin{figure}
\includegraphics[scale=0.9]{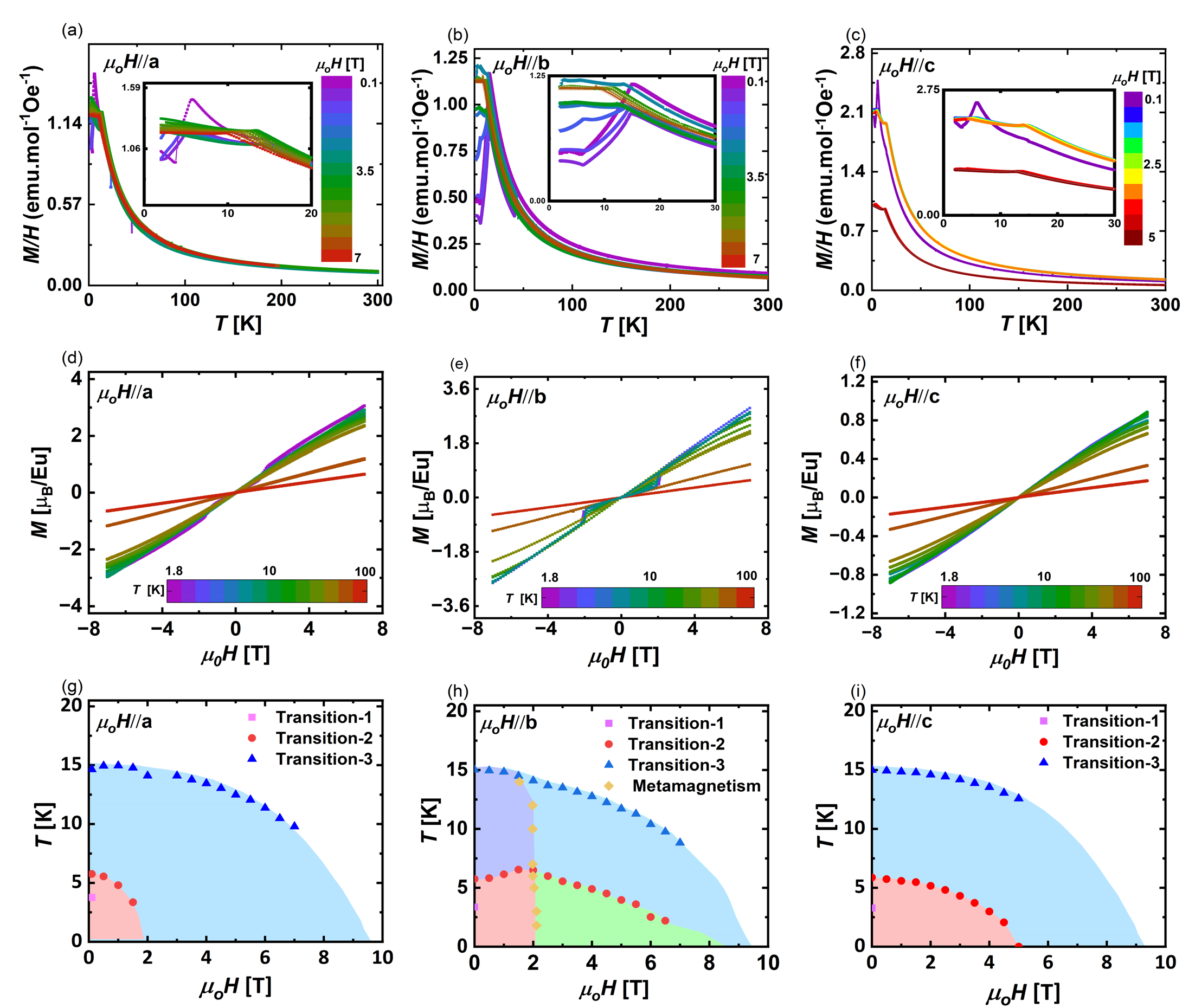}
\caption{\label{magnetism}
Magnetization as a function of temperature with  $\mu{o}\textit{H}$ = 0.1-7 T and T = 2-300 K in $\frac{2}{3}$ Eu\textsubscript{5}In\textsubscript{2-x}Ga\textsubscript{x}Sb\textsubscript6 single crystals at various applied magnetic fields at (a) $\mu{o}\textit{H}$//a, (b) $\mu{o}\textit{H}$//b, and (c) $\mu{o}\textit{H}$//c. There are two transitions at T=7.2K and T=14.4 K. Magnetization as a function of the magnetic field from $\mu{o}\textit{H}$=-7 to 7 T for (c) $\mu{o}\textit{H}$//a, (d) $\mu{o}\textit{H}$//b, and (e) $\mu{o}\textit{H}$//c. There is metamagnetism observed in $\mu{o}\textit{H}$//a and $\mu{o}\textit{H}$//b. Magnetic phase diagram (g) $\mu{o}\textit{H}$//a, (h) $\mu{o}\textit{H}$//b, and (i) $\mu{o}\textit{H}$//c.}\end{figure}
\pagebreak

\begin{figure}
\includegraphics[scale=1]{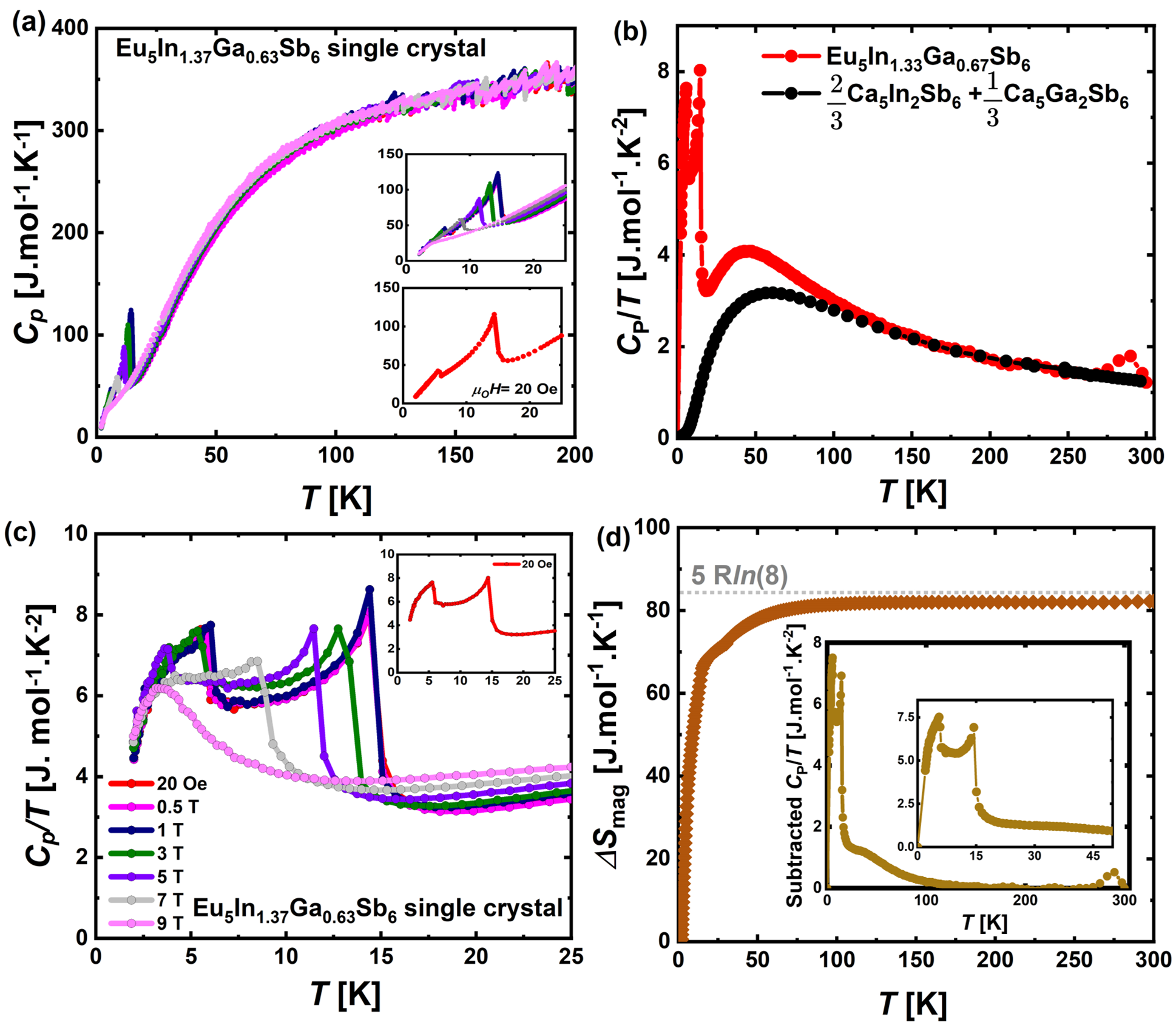}
\caption{\label{hc}
Heat capacity as a function of temperature for Eu\textsubscript{5}In\textsubscript{2-x}Ga\textsubscript{x}Sb\textsubscript{6} with  x=$\frac{2}{3}$ single crystals at various applied magnetic fields. There are two transitions at $T$=7.2K and $T$=14.4 K. (b) The $C_p/T$  plots for plot (a). (c) The phonon subtraction using the experimentally measured Ca$_5$In$_2$Sb$_6$ and Ca$_5$Ga$_2$Sb$_6$. The upturn at around $T$$\sim$280 K is because of Apezon N grease. (d) The integrated change in magnetic entropy for x=$\frac{2}{3}$ Eu\textsubscript{5}In\textsubscript{2-x}Ga\textsubscript{x}Sb\textsubscript6 single crystals that saturates at the theoretical 5 $R$ln(8) for a $J=S$=7/2 system.}
\end{figure}
\pagebreak
\begin{figure}
\includegraphics[scale=0.80]{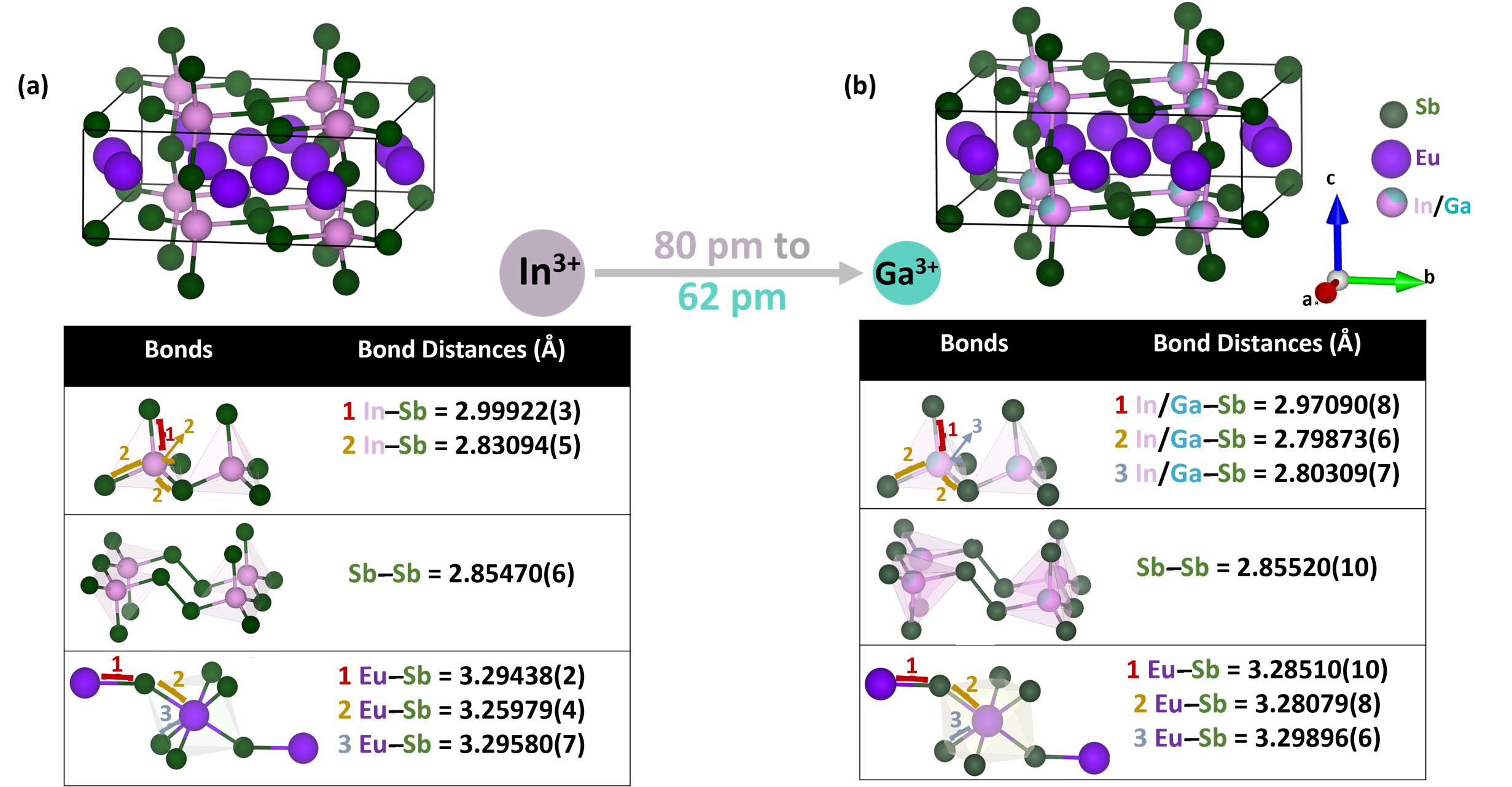}
\caption{\label{Fig-2.png}
Shift structural shifts from (a) Eu\textsubscript{5}In\textsubscript{2}Sb\textsubscript{6} to (b) x=$\frac{2}{3}$ Eu\textsubscript{5}In\textsubscript{2-x}Ga\textsubscript{x}Sb\textsubscript{6}. The tables below (a) and (b) explain the distortion in the trigonal pyramidal to a bent geometry in the (InSb$_4$) frameworks and the overall shortening of the bond distances on Ga substitution with more significant figures.}
\label{struc2}
\end{figure}

\pagebreak
\begin{figure}
\includegraphics[scale=4]{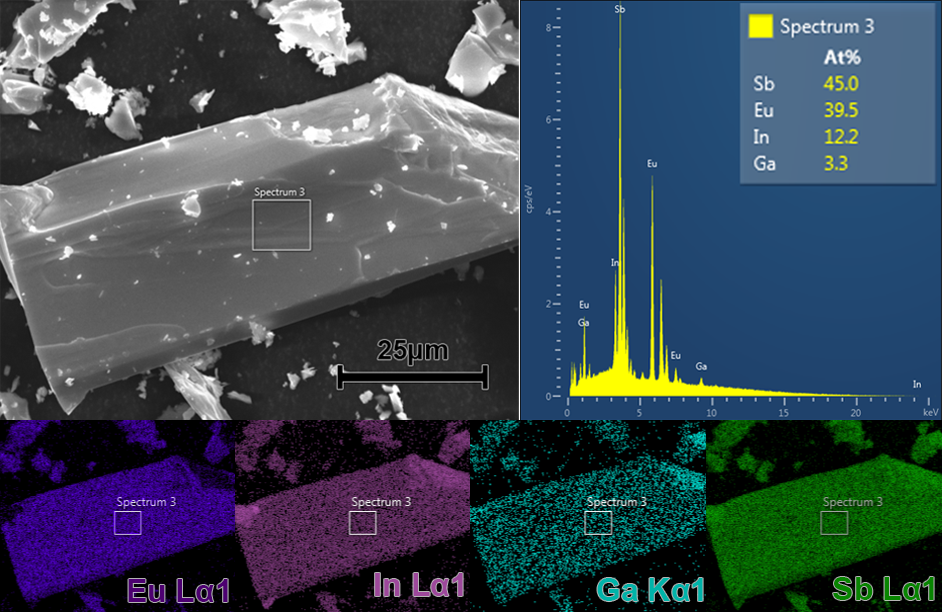}
\caption{\label{fin329.png}}SEM-EDS maps and spectrum of Sample-2 with the composition of Eu\textsubscript{5}In\textsubscript{1.586}Ga\textsubscript{0.429}Sb\textsubscript{6}.
\end{figure}
\pagebreak

\pagebreak
\begin{figure}
\includegraphics[scale=4]{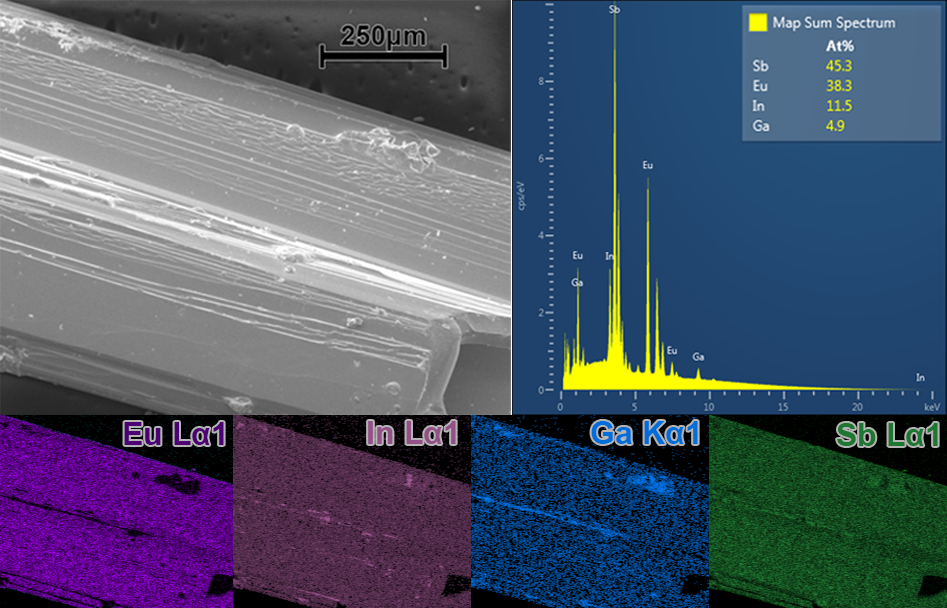}
\caption{\label{fin328.png}}SEM-EDS maps and spectrum of Sample-1 with the composition of Eu\textsubscript{5}In\textsubscript{1.495}Ga\textsubscript{0.637}Sb\textsubscript{6}.
\end{figure}
\pagebreak

\pagebreak
\begin{figure}
\includegraphics[scale=4]{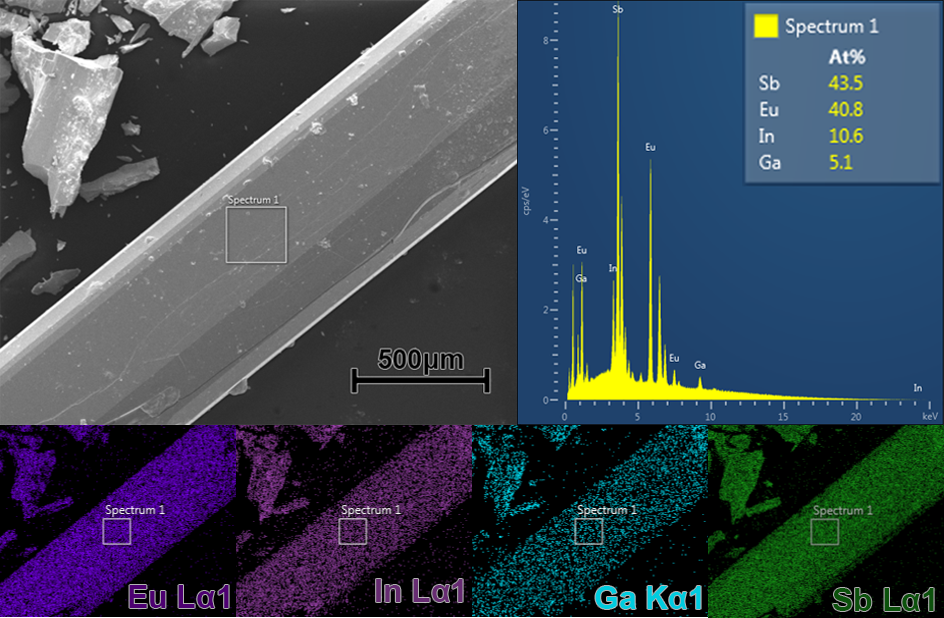}
\caption{\label{fin332.png}}SEM-EDS maps and spectrum of Sample-3 with the composition of Eu\textsubscript{5}In\textsubscript{1.378}Ga\textsubscript{0.663}Sb\textsubscript{6}.
\end{figure}
\pagebreak

\pagebreak
\begin{figure}
\includegraphics[scale=0.9]{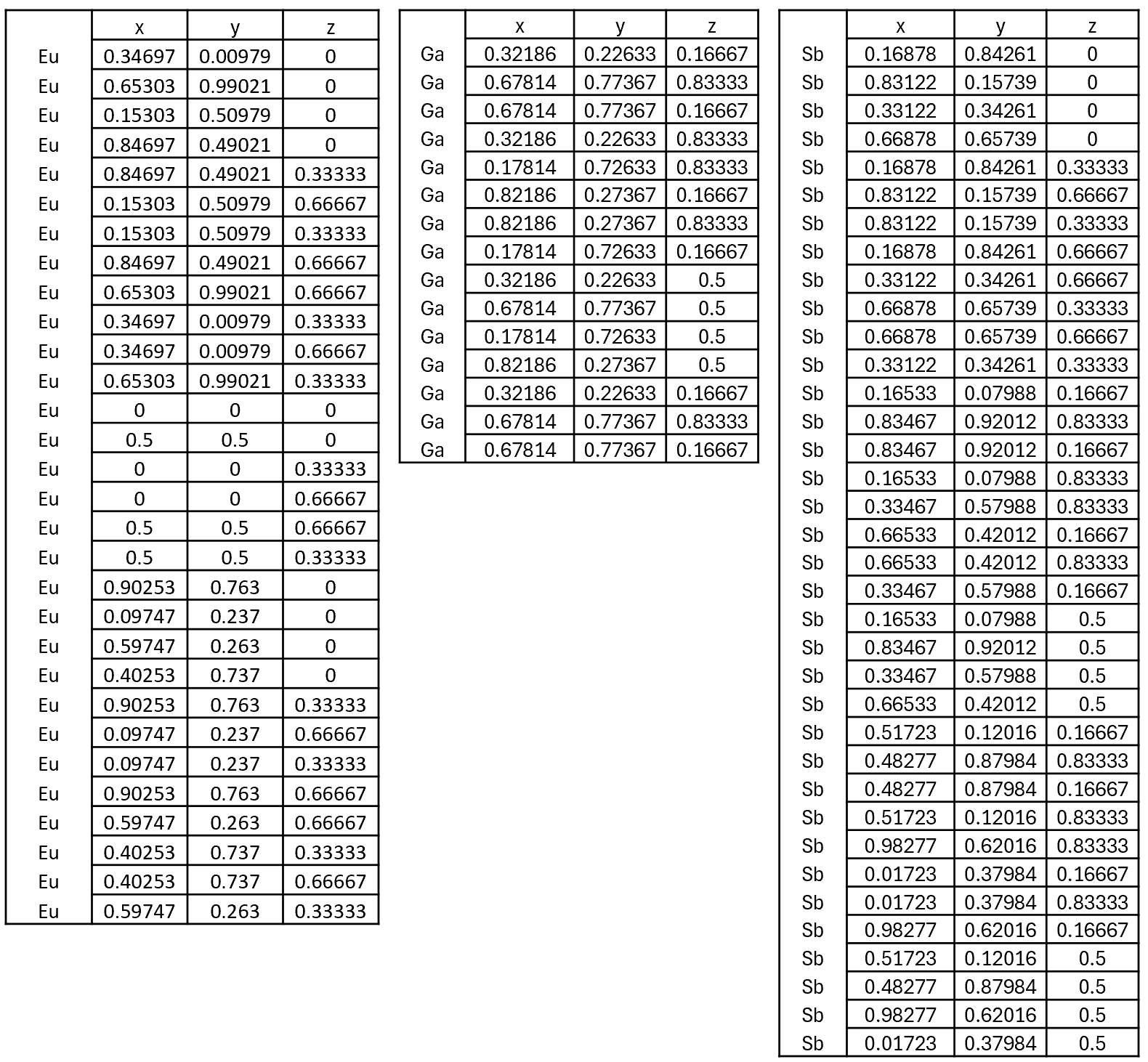}
\caption{\label{0Gasupercellposition.png}The DFT Supercell positions of x=2 in Eu\textsubscript{5}In\textsubscript{2-x}Ga\textsubscript{x}Sb\textsubscript{6}.}
\end{figure}
\pagebreak

\pagebreak
\begin{figure}
\includegraphics[scale=0.9]{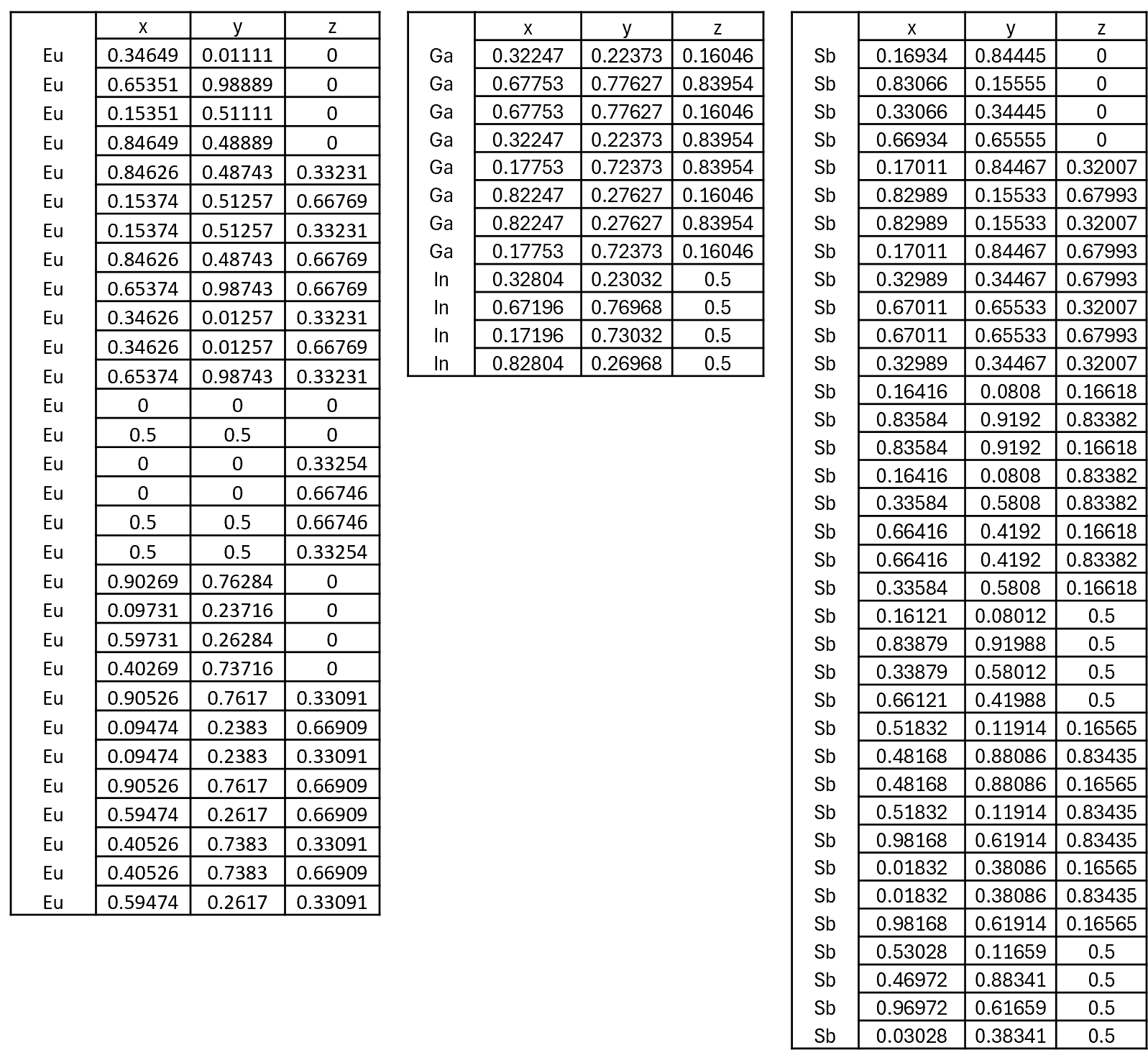}
\caption{\label{33Gasupercellposition.png}The DFT Supercell positions of x=4/3 in Eu\textsubscript{5}In\textsubscript{2-x}Ga\textsubscript{x}Sb\textsubscript{6}.}
\end{figure}
\pagebreak

\pagebreak
\begin{figure}
\includegraphics[scale=0.9]{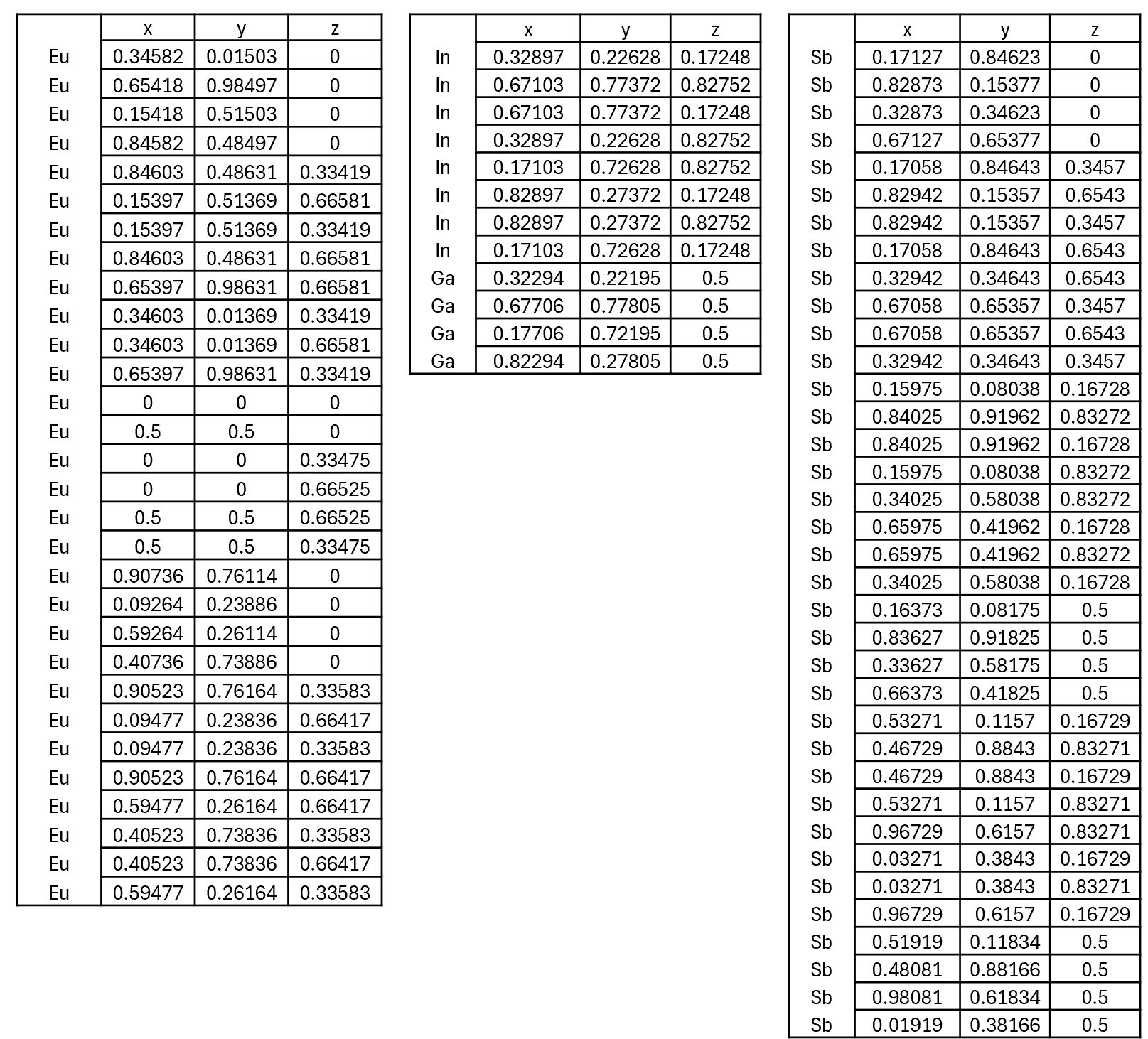}
\caption{\label{66Gasupercellposition.png}The DFT Supercell positions of x=2/3 in Eu\textsubscript{5}In\textsubscript{2-x}Ga\textsubscript{x}Sb\textsubscript{6}.}
\end{figure}
\pagebreak
\pagebreak

\begin{figure}
\includegraphics[scale=0.9]{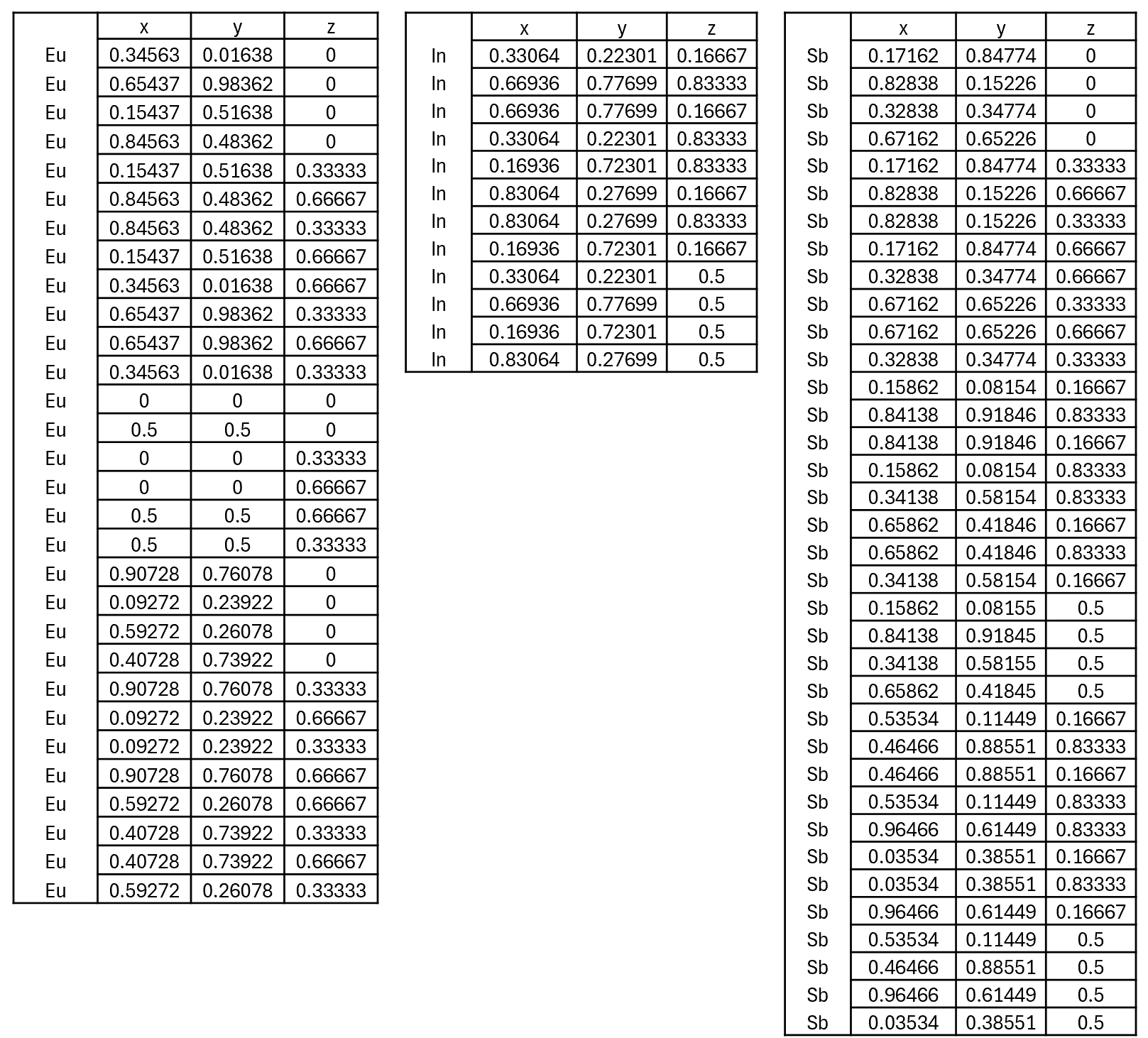}
\caption{\label{100Gasupercellposition.png}The DFT Supercell positions of x=0 in Eu\textsubscript{5}In\textsubscript{2-x}Ga\textsubscript{x}Sb\textsubscript{6}.}
\end{figure}
\pagebreak

\pagebreak
\begin{figure}
\includegraphics[scale=0.6]{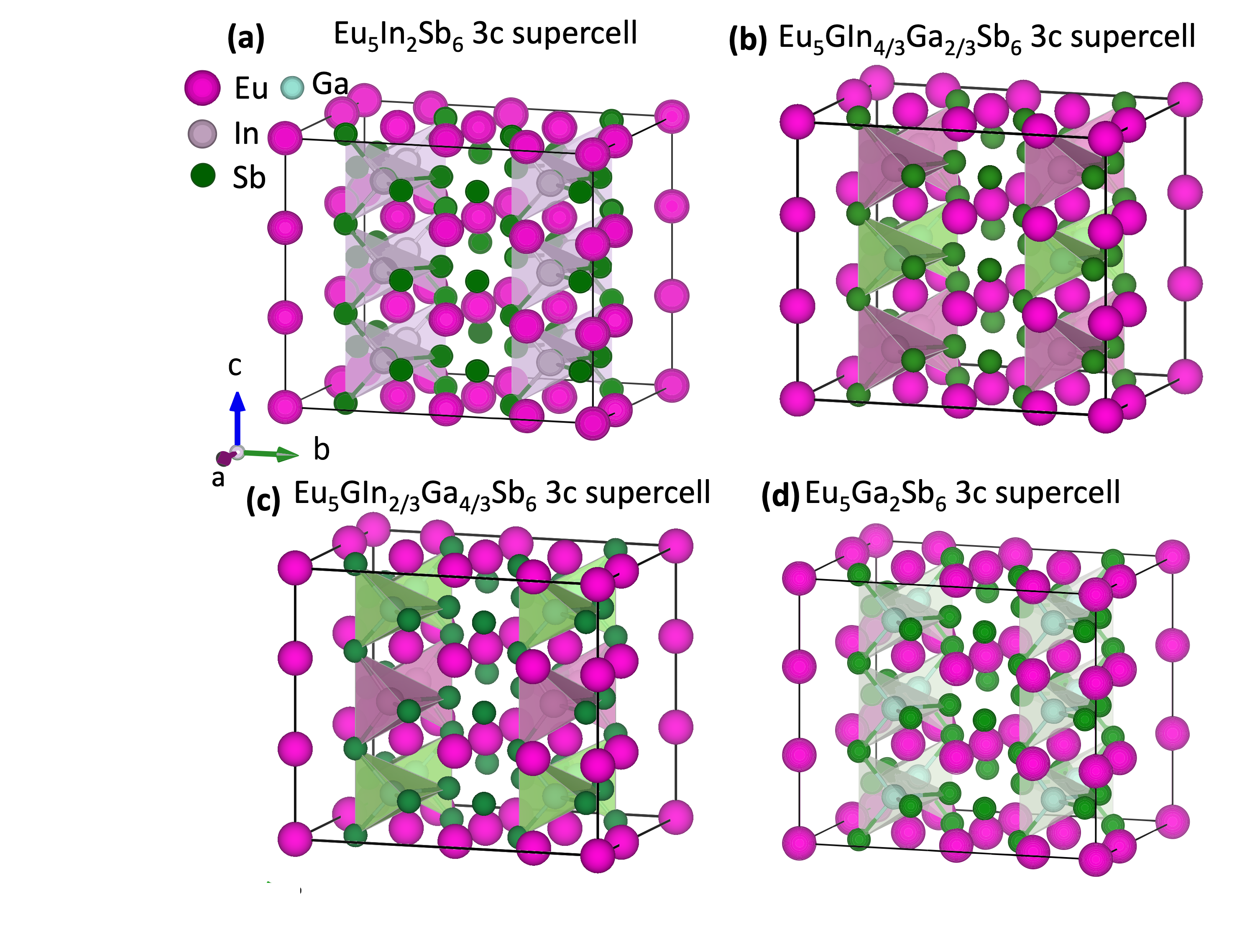}
\caption {The DFT Supercell structures for (a) x=0, (b) x=$\frac{2}{3}$, (c) x=$\frac{4}{3}$, and (d) x=2 in Eu\textsubscript{5}In\textsubscript{2-x}Ga\textsubscript{x}Sb\textsubscript{6} respectively.}
\end{figure}
\pagebreak
\pagebreak


\end{document}